\documentclass[pra,final,twocolumn,showpacs,floatfix,superscriptaddress,amsmath,amssymb,longbibliography]{revtex4-1}

\DeclareRobustCommand{\gobblefive}[5]{}
\newcommand*{\SkipTocEntry}{\addtocontents{toc}{\gobblefive}}

\usepackage{graphicx}
\usepackage{mathptmx}
\usepackage{color}
\definecolor{mygrey}{gray}{0.35}
\definecolor{myblue}{rgb}{0.2,0.2,0.8}
\definecolor{myzard}{cmyk}{0,0,0.05,0}
\definecolor{mywhite}{rgb}{1,1,1}
\definecolor{myred}{rgb}{1,0.,0.3}
\usepackage[colorlinks=true,citecolor=myblue,linkcolor=myred]{hyperref}

\usepackage{xcolor}
\usepackage[author={LR Team}]{pdfcomment}
\usepackage[english]{babel}
\usepackage[utf8]{inputenc} 
\usepackage{braket}
\usepackage{bbm}
\usepackage{listings}

\newcommand{\dd}{\mathrm{d}}

\newcommand{\vect}[1]{\mathbf{#1}}
\newcommand{\norm}[1]{\ensuremath{\left\Vert{#1}\right\Vert}}

\newcommand{\ii}{{\rm i}}
\newcommand{\ee}{{\rm e}}

\begin{document}
  
\title{Lieb-Robinson bounds for spin-boson lattice models and trapped ions}

\author{J. J\"unemann}
\affiliation{Instituto de F{\'\i}sica Fundamental IFF-CSIC, Calle Serrano 113b, Madrid E-28006, Spain}
\affiliation{Freie Universit\"at Berlin, Arnimallee 14, 14195 Berlin, Germany}

\author{A. Cadarso}
\affiliation{Instituto de F{\'\i}sica Fundamental IFF-CSIC, Calle Serrano 113b, Madrid E-28006, Spain}
\affiliation{Facultad de Matem\'aticas, Universidad Complutense de Madrid, Avenida Complutense s/n, Madrid E-28040, Spain}
\author{D. P\'erez-Garc{\'\i}a}
\affiliation{Facultad de Matem\'aticas, Universidad Complutense de Madrid, Avenida Complutense s/n, Madrid E-28040, Spain}
\author{A. Bermudez}
\affiliation{Instituto de F{\'\i}sica Fundamental IFF-CSIC, Calle Serrano 113b, Madrid E-28006, Spain}
\author{J. J. Garc{\'\i}a-Ripoll}
\affiliation{Instituto de F{\'\i}sica Fundamental IFF-CSIC, Calle Serrano 113b, Madrid E-28006, Spain}

\begin{abstract}
 We derive a Lieb-Robinson bound for the propagation of  spin correlations in a model of spins interacting through a bosonic lattice field, which satisfies itself a Lieb-Robinson bound in the absence of spin-boson couplings. We apply these bounds to a system of trapped ions, and find that the propagation of spin correlations, as mediated by the phonons of the ion crystal, can be  faster than the regimes currently explored in  experiments. We propose a scheme to test the bounds by measuring retarded  correlation functions via the crystal fluorescence.
\end{abstract}


\maketitle

The possibility of designing or simulating many-body systems in quantum-optical setups, such as ultracold atoms\ \cite{Endres2011,Cheneau2012a, Trotzky2012,Fukuhara2013} or large ion crystals\ \cite{Britton2012}, is stimulating considerable progress  in our understanding of non-equilibrium quantum many-body phenomena. However, the interpretation of these experiments demands powerful theoretical tools, including the {\it Lieb-Robinson bounds} (LRBs)\ \cite{Lieb1972} developed in this work. On the surface, LRBs show that non-relativistic quantum many-body systems, under certain conditions, display a causal structure analogous to relativistic quantum field theories. More deeply, LRBs are essential to prove fundamental  quantum many-body properties, such as the exponential decay of correlations in the ground-state of gapped local Hamiltonians ---the so-called ``clustering of correlations''\ \cite{Hastings2006}---, scaling laws for entanglement entropy\ \cite{Hastings2007e} ---the ``area laws''\ \cite{area_laws}
---, or the robustness of topological order under local perturbations\ \cite{top_order}.

{\it Causality} limits how local measurements and perturbations, described by an operator ${O}_X$ in region $X$, affect later measurements of another operator ${O}_Y$ in a separate region $Y$\ \cite{peskin}.  In analogy to Heisenberg's principle\ \cite{robertson}, this uncertainty is quantified by a commutator $C_{Y,X}(t)=\langle[O_Y(t),O_X(0)]\rangle$.  Lorentz invariance and the mathematical structure of relativistic theories guarantee causality\ \cite{peskin}. Thus, $C_{Y,X}(t)=0$ when the distance $d_{XY} > ct$ places both regions outside the light cone defined by the speed of light $c$. In non-relativistic quantum mechanics, causality is violated at the few particle level\ \cite{peskin}. Remarkably, in the many-body regime, an approximate light cone emerges, outside of which such correlations are vanishingly small. This phenomenon, first demonstrated by Lieb and Robinson\ \cite{Lieb1972} for a lattice of locally-interacting spins, has been generalized to finite-dimensional models, anharmonic oscillators 
and 
master equations\ \cite{Hastings2006,spin_LRB,Cramer2008c,Nachtergaele2008,Poulin2010b}. 

In this work, we address the role of bosons as mediators of interactions between particles in the light of LRBs. This is done for a general model of finite dimensional systems interacting through a bosonic field that satisfies a LRB itself.  New bounds are derived, which are then applied to a crystal of trapped atomic ions, where the spins and the bosons map to the ions' internal states and the crystal's phonons, respectively. These LRBs work for all {\it spin-boson lattice models} of any dimensionality and geometry realized with state-of-the-art technology\ \cite{Ising_1d,tunneling_1d,Britton2012,hensinger}.  Comparing with LRBs for the effective spin models\ \cite{Porras2004,Porras2006} in quantum simulations\ \cite{Ising_1d,Britton2012}, we show that correlations can spread much faster in the non-perturbative regime. We note that the correlation spread in the perturbative regime has received considerable attention\ \cite{correlations_ising_models} and also remark that our results immediately extend to a 
variety of other fields, such as superconducting quantum circuits and quantum dots or NV-centers interacting with coupled cavities or photonic crystals.

{\it The  model.--} We consider a lattice model of bosons interacting locally with a collection of  finite-dimensional quantum systems. The lattice is an undirected graph, where each of the $N$ vertices forms a Hilbert space that groups a boson with a system of finite dimension $d$, i.e. ``spin''. The Hamiltonian  is

\begin{equation}
  H= \frac{1}{2} \sum_{i,j=1}^N \vect{R}_i^T Q_{ij}(t) \vect{R}_j + \sum_{i=1}^N \vect{B}_i(t)^T \vect{S}_i+
  \sum_{i=1}^N \vect{R}_i^T  G_{i}(t)  \vect{S}_i.
  \label{eq:model}
\end{equation}
Here, the bosons are represented by adimensionalized harmonic oscillators with positions and momenta $\mathbf{R}^T_i=(x_i,p_i)$ satisfying $[\vect{R}_i, \vect{R}_j^T] = - \delta_{i,j} \sigma^y$, with the Kronecker delta $\delta_{i,j}$ and the Pauli matrix $\sigma^y$. The spins $\vect{S}_i=(S^1_i,\ldots,S^{m}_i)$ are dimensionless operators forming a Lie algebra with structure constants $f^{\alpha\beta\gamma}$~\cite{comment_lie_algebra}. 
The dynamics of this spin-boson lattice is given by Eq.\ \eqref{eq:model}, where bosons at different vertices are coupled by a matrix $Q_{ij}(t)=Q_{ji}(t)\in\mathbb{R}^{2\times2}$, and  spins precess under a general magnetic field $\vect{B}^T_i(t)=(B_i^1(t),B^2_i(t),B^3_i(t))$. Finally, the spin-boson coupling is strictly local, taking place exclusively at the vertices through matrices $G_i(t) \in \mathbb{R}^{2\times3}$.

Whereas the first part of this Letter introduces the proof of the LRB for the very general spin-boson Hamiltonian of Eq.~\eqref{eq:model}, the second part shows how this model applies to trapped ions. We estimate correlation speeds and timescales, suggesting concrete experimental protocols to assert these bounds. As an aid to the reader, in the Supplemental Material~\cite{supp_material} we provide additional background material, a step-by-step version of the proof, and alternative experimental setups or considerations.

{\it Spin-boson LRB.--}
Let us assume that the propagator $W$ of the free bosonic lattice without spins ($G_i=0$) satisfies a LRB

\begin{equation}
  \norm{[\vect{R}_j(t),\vect{R}_k(0)]} \leq \norm{W_{jk}(t,0)} \leq \, \alpha\, e^{\nu_{\rm LR}t}f(d_{jk}),
\label{eq:harmonic-bound}
\end{equation}

characterized by a LR speed $v_{\rm LR}$, a normalization $\alpha>0$, and a function of the lattice distance such that

\begin{equation}
  a_0 :=\max_{ik} \big[f(d_{ik})^{-1}\sum_j f(d_{ij}) f(d_{jk})\big] < +\infty.
  \label{a0}
\end{equation}

Under these conditions, assuming bounded interactions $\Vert{G_j(t)}\Vert\le g$ and spins $\Vert \vect{S}\Vert \leq S$, a LRB emerges for the spin correlations $\vect{Z}_{jk}(t) := [\mathbf{S}_j(t), S^\phi_k(0)]$, $\phi\in\{1,2,3\}$

\begin{align}
\Vert\mathbf{Z}_{jk}(t)\Vert  \le  \alpha\, e^{\nu_{\rm LR} t}f(d_{jk}) \times \frac{2S^2}{a_0}
\bigg(\ee^{(g^2/v_{\rm LR}) 2S\alpha a_0 t}-1\bigg).
\label{LR_bound}
\end{align}

Intuitively, since bosons mediate interactions, the bosonic velocity $v_{\mathrm{LR}}$ limits the propagation speed of spin correlations. This is precisely the first term of the above expression, which duplicates the bosonic LRB. Additionally, the efficiency with which distant spins excite and reabsorb a propagating boson affects the LRB. This is the second term in Eq.~\eqref{LR_bound}, which depends on the rate $\sim g^2/v_{\rm LR}$ at which bosons are emitted or absorbed by spins. This nonperturbative correction shows that the buildup of correlations is suppressed if bosons are much faster than spins $g\ll v_{\rm LR}$, an adiabatic-type argument.

Note that Eqs.\ (\ref{eq:harmonic-bound}) and (\ref{a0}) include a very large family of bounds

\begin{align}
e^{\nu_{\rm LR} t} f(d_{jk}) = e^{\nu_{\rm LR} t - \mu d_{jk}} (1+ d_{jk})^{-\eta},
\label{LRB_boson_local}
\end{align}

with appropriate $\mu \ge 0$ and $\eta>0$. For nearest-neighbor or short-range interactions $\mu > 0$ yields a light cone, $\mu d_{jk} - v_{\rm LR}t \sim 0$, outside of which correlations are exponentially suppressed. For algebraically decaying interactions, $\mu = 0$, and the lines of constant correlation are only straight at short distances and times.

{\it The proof.--} We will sketch the technical steps to recover this result (see the Supplemental Material\ \cite{supp_material}). The Heisenberg equations of motion are $\dot{\vect{R}}_j = - \sum_{k} J Q_{jk}(t) \vect{R}_k - J G_j(t) \vect{S}_{j}$ and $\dot{\vect{S}}_j = \ii K_{j}(t) \mathbf{S}_j$, with a Hermitian matrix $K_j(t)$ that depends on the couplings, boson operators, and spin structure constants. The first equation is formally integrated $\mathbf{R}_j(t) = \sum_kW_{jk} (t,0)\mathbf{R}_k(0) - \int_0^t \mathrm{d}\tau \sum_kW_{jk}(t,\tau) J G_k(\tau) \vect{S}_k(\tau)$. In this notation, the free boson propagator $W$ is an $N\times N$ block matrix, where each block $W_{jk}(t_1,t_2)\in\mathbb{R}^{2\times 2}$ spreads correlations between sites $j$ and $k$.

The bosonic bound\ \eqref{eq:harmonic-bound} influences the spin-spin correlations through the spin-boson correlators $\vect{C}_{jk}(t) := [\mathbf{R}_j(t), S^\phi_k(0)]$. This is seen in $ \dot{\vect{Z}}_{jk} = i \sum_{n\in\{x,p\}} C_{jk}^n A^n_j(t) \vect{S}_j +\ii K_j(t) \vect{Z}_{jk}$, where  the matrices $A^n_j(t)$ are defined in terms of the spin-boson couplings (Supplemental Material\ \cite{supp_material}).  To eliminate the local precession of the spins, we change variables $\mathbf{D}_{jk}(t) := O^{-1}_j(t) \mathbf{Z}_{jk}(t)$ with a unitary $O_j(t)$ obtained by solving $ \dot{O}_j(t) = i K_j(t) O_j(t)$. Thus,%

\begin{align}
  \dot{\vect{C}}_{jk} &= - \sum_{l} J  Q_{jl}(t) \vect{C}_{lk} - J  G_j(t)  O_j(t) \vect{D}_{jk}\label{eq:C},\\
  \dot{\vect{D}}_{jk} &= i O_j^{-1}(t)  \sum_n C_{jk}^n \, A^n_j(t)   \vect{S}_j \label{eq:D}
\end{align}

describe the buildup of spin-boson correlations\ \eqref{eq:C} and the conversion of spin-boson into spin-spin correlations\ \eqref{eq:D}. Some remarks are in order: (i) the equation for $\vect{C}_{jk}$ is solved formally in terms of $\vect{D}_{jk}$, creating a recursion; (ii) the operator $O_i$ absorbing the unbounded local rotations does not influence the LRB because (iii) $\mathbf{D}_{jk}$ and $\mathbf{Z}_{jk}$ have the same operator norms.

Equations\ \eqref{eq:C} and\ \eqref{eq:D}, with the upper bounds $\Vert{G_j(t)}\Vert,\Vert{A^n_j(t)}\Vert\leq g$, $\Vert{\vect{S}_j(t)}\Vert\leq S$, the bosonic LRB~\eqref{eq:harmonic-bound}, and the geometric factor $a_0$, provide a Dyson-type recursion for the commutators norms%

\begin{align}
  &\Vert\mathbf{D}_{jk}(t)\Vert \le \Vert{\mathbf{D}}_{jk}(0)\Vert\!+\\
&  \quad\quad + 2 g^2 S \alpha\sum_{l} \int_0^t \!\!\!\!d\tau_1\!\!
  \int_0^{\tau_1}\!\!\!\!\!\!d\tau_2 f(d_{jl},v_{\rm LR}(\tau_1-\tau_2)) \Vert\mathbf{D}_{lk}(\tau_2)\Vert.
 \nonumber
\end{align}

After summing this recursion to infinite order, the desired LRB\ \eqref{LR_bound} for the spin-boson lattice is recovered. $\square$

{\it Bosonic LRB.--}
 For the previous results to be useful, the bound of Eq.~\eqref{eq:harmonic-bound} must be sufficiently tight.  Whereas the bosonic LR speed has been studied for nearest-neighbor\ \cite{Nachtergaele2008} and algebraically decaying\ \cite{Cramer2008c} time-independent couplings $Q$, we have developed a tighter bound for such models (Supplemental Material\ \cite{supp_material}). Our LR speed only relies on the off-diagonal couplings, using a recursion similar to that in Ref.\ \cite{Cramer2008c}, but eliminating the diagonal terms with unitary transformations. The case relevant for trapped ions involves long-range interactions

\begin{align}
  \norm{Q_{jk}(t)}_{j\neq k}\leq\kappa (1+d_{jk})^{-\eta}
  \label{eq:power-law}
\end{align}

with a strength $\kappa$ that bounds the couplings and a decay power $\eta\geq 0$. We then recover Eq.~\eqref{eq:harmonic-bound} with $f(d)=(1+d)^{-\eta}$, $\alpha = (1+a_0)/a_0$, and a bosonic LR speed $v_{\rm LR}=\kappa a_0$. 

{\it Trapped-ion implementation.--}
Equation\ \eqref{LR_bound} applies to a great variety of systems. In particular we show below that the case of trapped ions, a prominent architecture for quantum information\ \cite{haeffner_review}, is an ideal setup to experimentally test our LRB.

Laser-cooled ions in radio-frequency or Penning traps form crystals of tunable dimensionality and geometry. The spin-boson model of Eq.~\eqref{eq:model} can be implemented on top of this state-of-the-art technology. The {\it bosonic lattice} describes the small transverse displacements $\vect{R}$ of the ions from their equilibrium positions in the crystal. The {\it spins} $\vect{S}=\vec{\sigma}$ are encoded in two levels of the atomic structure $\{\ket{{\uparrow}},\ket{{\downarrow}}\}$ with long coherence times\ \cite{comment_ions}. The {\it spin-boson coupling} is provided by dipole forces that push the atoms depending on their internal state\ \cite{sd_forces}. Without loss of generality\ \cite{sd_forces_note}, we assume that near-field microwaves or lasers induce a uniform force in the $\sigma^z$-basis. In Eq.~\eqref{eq:model} this maps to $G_i^{n\gamma}(t)=F_z(t) \delta_{n,x}\delta_{\gamma,3}$, where the strength $F_z=\sqrt{2}\tilde{\Omega}\gamma$ depends on the field intensity or Rabi frequency $\tilde{\
Omega}(t)
$ and the Lamb-Dicke parameter $\gamma\ll1$.

For this trapped-ion implementation of the spin-boson lattice, we can evaluate the LRB\ \eqref{LR_bound}, obtaining a power-law behavior\ \eqref{LRB_boson_local}. We start by considering the bosonic part of the evolution. The phonon coupling $Q_{ij}={\rm diag}\{\omega_{\rm t}\delta_{i,j}+\mathbb{V}_{ij}(m\omega_{\rm t})^{-1},\,\,\omega_{\rm t}\delta_{i,j} \}$ contains the ions' mass $m$, the transverse trap frequency $\omega_t$, and a dipolar interaction\ \cite{coulomb_couplings}. The off-diagonal couplings thus satisfy Eq.~\eqref{eq:power-law} with algebraic decay $\eta=3$, $d_{ij}$ being the lattice distance of the ions in the crystal. The interaction strength $\kappa =4\beta \omega_t$ is defined in terms of the stiffness parameter\ \cite{Porras2004}, $\beta=e^2/4\pi\epsilon_0m\omega_{t}^2d_{m}^3$, which measures the ratio of the Coulomb repulsion to the trapping energy and depends on the minimal separation between two ions in the crystal, $d_{m}$. Introducing the maximum force $g=\max_t |F_z(
t)|$ and $S=1$, the bound reads

\begin{equation}
\Vert [\boldsymbol{\sigma}_i(t), \sigma^{\phi}_j(0)] \Vert_{\infty}\leq {2  \over a_0 (1 + d_{ij})^3} e^{\alpha_1(\beta\omega_{t})t}\big( e^{\alpha_2(\beta\omega_{t})t} -1 \big),
\label{lrb_ions_spin_boson}
\end{equation}%

where we  define $\alpha_1=8a_0$ and $\alpha_2=(1/4)(1+a_0^{-1})(g/ \beta\omega_{\rm t})^2$. As anticipated in the proof, the LRB depends fundamentally on the maximum group velocity of the phonon branch, given by $\beta\omega_{\rm t}$, and on the efficiency of the force in exciting and absorbing a propagating phonon, $(g^2/ \beta\omega_{t})$.

\begin{figure}[t]
  \centering
  \includegraphics[width=\linewidth]{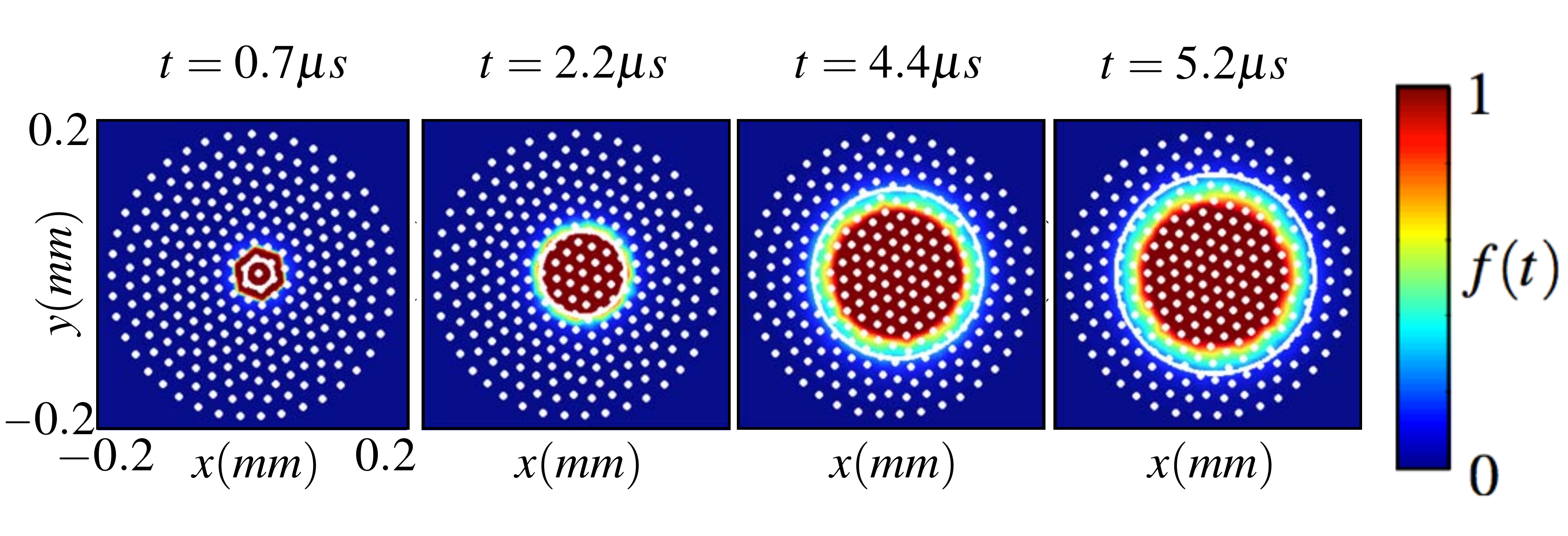}
  \caption{{\bf Spin correlation spread in the impulsive regime.} In this regime, we evaluate numerically the bound $\Vert [{\sigma}^x_i(t), \sigma^{x}_j(0)] \Vert_{\infty}\leq f(t)$, where $f(t)={\rm max}_{\tau\leq t}\{8|\sin(W_{ij}^{xp}(\tau,0))|\}\times\theta_i\theta_j$ is obtained from the exact time-evolution of the impulsive regime. We consider a crystal of $N=253$ $^9$Be$^+$ ions\ \cite{Britton2012}, assuming pulse areas of $\theta_l=1$. The white circle corresponds to a wavefront advancing at a speed of $3 d_m \beta \omega_t$.}
  \label{fig:ions}
\end{figure}

To evaluate Eq.~\eqref{lrb_ions_spin_boson} in a realistic experimental situation, we  focus on $^9{\rm Be}^+$ ions in a Penning trap\ \cite{Britton2012} and refer the reader to the Supplemental Material\ \cite{supp_material}  for other  setups. These ions, confined with a transverse trap frequency of $\omega_{t}/2\pi\approx0.8\,$MHz, form a triangular crystal of $N\sim$100-300 lattice sites  characterized by a minimal  distance $d_{m}\sim20\,\mu$m. In such experiments, the maximum phonon group velocity is currently $\beta\omega_{t}/2\pi\approx 60\,$kHz, and oscillating state-dependent forces with $g/2\pi\approx0.6\,$kHz have been obtained from two non-copropagating laser beams in a Raman configuration. As discussed in the Supplemental Material\ \cite{supp_material},  by employing larger  angles of the incident beams, and short pulses that relieve the need for compensating ac-Stark shifts and resolving sidebands, the strength of the forces can be increased to $g/2\pi\approx0.3\,$MHz. Still higher forces can 
be achieved with counter-propagating pairs 
of ultra-fast laser pulses\ \cite{fast_forces}. 

The last piece of information to evaluate\ \eqref{lrb_ions_spin_boson} is the  parameter $a_0$. A crude approximation is to use an infinite lattice with uniform geometry (triangular or one-dimensional) and obtain $a_0$ from the convolution defined above\ \eqref{a0}. This leads to $a_0\approx 8.5$ and would set the correlation timescale at $\sim 0.1-1 \mu$s.



{\it Impulsive regime.--} We will now discuss a regime  where the correlation timescale can approach the optimal prediction  of the LRB\ \eqref{lrb_ions_spin_boson}. The { impulsive regime} relies on strong forces $g\gg\beta\omega_{t}$  during a short lapse $\delta t\sim g^{-1}$. In particular, we assume that a pulsed force is applied at $t=0$ to the $j$-th ion to create a bosonic excitation correlated to the spin. After propagation of the bosonic field, another pulsed force is applied at the distant $i$-th ion to create the spin-spin correlations. The evolution operator is obtained analytically (Supplemental Material\ \cite{supp_material}), without approximations

\begin{equation}
\Vert [{\sigma}^x_i(t), \sigma^{x}_j(0)] \Vert_{\infty}\leq {8(1+a_0)  \over a_0 (1 + d_{ij})^3} e^{\alpha_1(\beta\omega_{t})t}\times|\theta_i\theta_j|,
\label{LRB_impulsive}
\end{equation}
 where $\alpha_1=8a_0$. The pulse area $\theta_l=\int_{0}^{ t} d \tau F_{z,l}(\tau)$ gives a measure of the number of phonons excited $\bar{n}_{t}\sim|\theta_l|^2$. As argued above, forces can be as large as $g/2\pi\approx0.3\,$MHz for $^9{\rm Be}^+$ ions in a Penning trap so that $\beta\omega_{\rm t}/g\approx 0.2$ and we achieve the impulsive regime. Note that the pulsed forces should be switched on/off in $\delta t\sim$0.1-1$\,\mu$s. In Fig.\ \ref{fig:ions}, we numerically solve the impulsive time-evolution (see Supplemental Material\ \cite{supp_material}). The spread of correlations is much faster ($\,\sim10\,\mu$s) than experimental decoherence (without reaching the theoretical maximum).


{\it Perturbative regime.--} Whereas our LRB\ \eqref{lrb_ions_spin_boson} gives the fastest timescale of correlation, many experiments for the simulation of quantum magnetism\ \cite{Ising_1d,Britton2012} are implemented in the so-called {perturbative regime}, which leads to significantly slower correlation speeds. The perturbative regime is characterized by oscillating forces that are much weaker than their detuning from the trap frequency, $g\ll{\delta}_{t}$. In this regime, phonons can be approximately traced out, leading to effective algebraically decaying spin-spin interactions\ \cite{Porras2004,Porras2006}, $H_{\rm eff}=\sum_{ij}J_{ij}\sigma_i^z\sigma_j^z$. When $\beta\omega_{t}\ll 2\delta_{t}$ , the effective interaction is dipolar $\Vert{J_{ij}}\Vert\leq 8J_0/(1+d_{ij})^3$, with $J_0=(g/4{\delta}_{t})^2\beta\omega_{t}$. In this case, we can apply the existing LRBs for spin models\ \cite{spin_LRB}, obtaining

\begin{equation}
\Vert [\boldsymbol{\sigma}_i(t), \sigma^{\phi}_j(0)] \Vert_{\infty}\leq {2  \over a_0 (1 + d_{ij})^3} \big( e^{\tilde{\alpha}_2(\beta\omega_{t})t} -1 \big),
\label{lrb_ions_spin_spin}
\end{equation}
 
where $\tilde{\alpha}_2=\frac{a_0}{8}(g/ \delta_{t})^2$. For the large detunings required to obtain the dipolar decay, $\delta_{t}/2\pi \approx 80\,$kHz, the propagation of spin correlations\ \eqref{lrb_ions_spin_spin} becomes very slow ($\sim1\,$s). For this reason, experiments use smaller detunings\ \cite{Britton2012}, leading to stronger couplings  at the expense of becoming truly long-ranged where the paradigm of LRBs no longer applies\ \cite{spin_LRB}. Yet, one can still expect correlation propagation in timescales$\,\sim 1\,$ms from the effective  model, which are still much slower than the optimal LRB\ \eqref{lrb_ions_spin_boson}\ \cite{comment_temperature}.

\begin{figure}
\centering
\includegraphics[width=1\columnwidth]{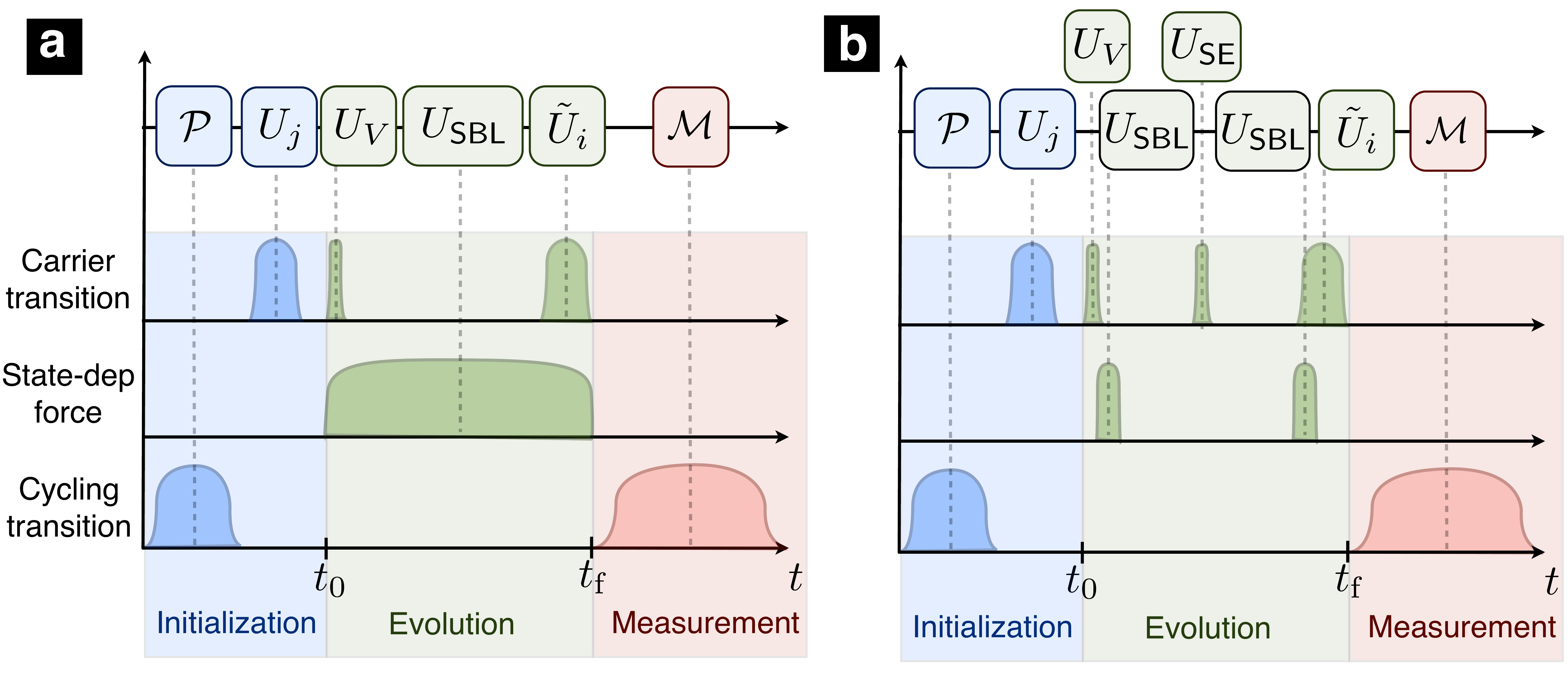}
\caption{ {\bf Experimental sequence to test the LRB: }  {\bf (a)} Always-on and {\bf (b)} pulsed spin-phonon forces. We represent  the {\it initialization} step in blue, which consists of laser cooling followed by optical pumping $\mathcal{P}$ and leads to $\ket{{\downarrow\cdots\downarrow}}\bra{{\downarrow\cdots\downarrow}}\otimes\rho_{\rm th}$, where $\rho_{\rm th}$ is a thermal state of the phonons after Doppler cooling. We then apply a $\pi/2$-pulse $U_{j}={\rm exp}\{\ii\frac{\pi}{2}\sigma_j^y\}$ by driving the carrier transition\ \cite{haeffner_review}. In the {\it measurement} step in red, one collects  the state-dependent fluorescence $\mathcal{M}$ during a continuous driving of the cycling transition\ \cite{haeffner_review}. At the beginning of the {\it evolution} step $t=t_0$, we  apply the unitary $U_{V}$ associated to the impulsive perturbation $V(t)$ described in the main text. This is followed by the actual evolution under the state-dependent forces: {\bf (a)} in the always-on regime, the 
forces should be switched on continuously during the evolution, or {\bf (b)} in the impulsive regime, we apply two pulsed  forces. Additionally, at the middle of the evolution, we apply the spin-echo sequence $U_{\textsf{SE}}$ $\sigma_i^z\to-\sigma_i^z$ and $\tilde{\Omega}\to-\tilde{\Omega}$ to refocus uncompensated ac-Stark shifts. Before measuring, we apply   another $\pi/2$-pulse $\tilde{U}_i={\rm exp}\{-\ii\frac{\pi}{2}\sigma_i^y\}$. }
\label{fig_LRB_measurement}
\end{figure}

{\it Probing the LRB through fluorescence.--} We discuss how to exploit the  control and measurement tools of trapped-ion experiments\ \cite{haeffner_review}  to probe  LRBs. Note that single-time observables, e.g. $\langle\sigma_i^\alpha(t)\rangle, \langle\sigma_i^\alpha(t)\sigma_j^\beta(t)\rangle$, are already being  measured with trapped ions\ \cite{state_tomography} or atoms in optical lattices\ \cite{Endres2011,dyn_correl}. Our aim is to measure retarded correlation functions  $\langle \sigma_i^{\alpha}(t)\sigma_j^{\beta}(0)\rangle$.

The experimental scheme to accomplish it, both for always-on (Fig.\ \ref{fig_LRB_measurement}{\bf (a)}) and pulsed  (Fig.\ \ref{fig_LRB_measurement}{\bf (b)})  state-dependent forces, is composed of three steps: {\it (i)} The {\it initialization} consists of preparing a localized spin excitation  $\ket{+}=(\ket{{\uparrow}}+\ket{{\downarrow}})/\sqrt{2}$ by a $\pi/2$-pulse at the $j$-th ion\ \cite{addressability}, while the phonon lattice is in  a thermal  state $\rho_{\rm th}$ (details in Fig.\ \ref{fig_LRB_measurement}),  such that $\rho_0=\ket{{\downarrow\cdots+_j\cdots\downarrow}}\bra{{\downarrow\cdots+_j\cdots\downarrow}}\otimes\rho_{\rm th}$. {\it (ii)} The {\it evolution} consists of letting the excitation propagate for $t\in(t_0,t_{\rm f})$, while switching on the state-dependent forces continuously (Fig.\ \ref{fig_LRB_measurement}{\bf (a)}) or in two pulses (Fig.\ \ref{fig_LRB_measurement}{\bf (b)}). To measure the retarded spin correlation functions to test  the LRBs, we need to apply two  unitaries 
in addition to the forces. First, we should apply an impulsive perturbation  $V(t)=\lambda_{B}\sigma_j^x\delta(t-t_0)$ localized at the $j$-th ion, and $\lambda_{B}\ll 1$. Second, after letting the system evolve, we should apply a  $\pi/2$-pulse at the distant $i$-th ion. {\it (iii)} The {\it measurement} consists of collecting the state-dependent fluorescence  of the ion crystal, which amounts to a measurement of $\langle\sigma_i^z(t_{\rm f})\rangle$. Using a linear-response-theory-type argument (Supplemental Material\ \cite{supp_material}), we have shown that $\partial_{\lambda_{B}}\langle\sigma_i^z(t_{\rm f})\rangle|_{\lambda_{B}=0}=-\ii\langle \sigma_i^x(t_{\rm f})\sigma_j^x(t_0)\rangle$. By modification of the unitaries\ \cite{supp_material,comm_position}, it is possible to  recover any retarded spin correlation $\langle \sigma_i^{\alpha}(t_{\rm f})\sigma_j^{\beta}(t_0)\rangle$. We remark that the operations required in each step are performed individually with accuracies better than $99\%$ in current 
experiments\ \cite{haeffner_review}.

{\it Conclusions.--} We have derived a new LRB for a collection of models\ \eqref{eq:model} involving spins and bosons in a lattice. Although the LRB applies to a variety of quantum-optical setups (e.g. superconducting circuits), we have focused on ion crystals, where it pinpoints that spin correlations can spread much faster than the experimental regimes currently considered. Regardless of the infinite dimensionality of Eq.~\eqref{eq:model}, we conjecture that the LRB encloses further theoretical implications, such as the clustering of correlations or the efficiency of time-dependent density matrix renormalization group methods, which might be of interest to recent studies\ \cite{sblm}.

{\it Acknowledgments.--} This work was supported by the European project PROMISCE, CAM research consortium QUITEMAD (S2009-ESP-1594), and the Spanish MINECO Project FIS2012-33022.

{\it Note added.--} Upon completion of this manuscript, we became aware of the e-print of Ref.~\cite{dyn_correl}, figuring a similar measurement protocol for spin correlations.

\SkipTocEntry

\clearpage

\begin{widetext}

\SkipTocEntry\section*{Supplemental material to ``Lieb-Robinson bounds for spin-boson lattice models and trapped ions''}

\begingroup
\hypersetup{linkcolor=black}
\tableofcontents
\endgroup

\section{Extended proof of the Lieb-Robinson bounds for  spin-boson models}
\label{LRB_spin_boson_sec}

In this section, we provide a detailed proof of the Lieb-Robinson bound (LRB) for the spin-boson lattice models.

\subsection{Statement of the problem}
\label{sec:statement}

Let us start by defining the system under study, which we refer to as the {\it spin-boson lattice model} (SBL). The lattice will be described by an undirected graph $G=(L,E)$ with a set of vertices $L$ where the physical degrees of freedom are defined, and an edge set $E,$ which describes neighbourhood relations in the lattice. The physical degrees of freedom of each vertex $i\in L$ are defined in the Hilbert space $\tilde{\mathcal{H}}_i = \mathcal{L}^2(\mathbb{R}) \otimes \mathcal{H}_i,$ which combines an infinite-dimensional  Hilbert space for the bosons, $\mathcal{L}^2(\mathbb{R})$, with a Hilbert space $\mathcal{H}_i$ of finite dimension $d_i$ for the  ``spins'' (see Fig.~\ref{SBLM_scheme}).

The bosons are represented by harmonic oscillators, whose positions and momenta  are grouped into a vector 
\begin{equation}
  \vect{R}^T = (\vect{R}_1^T,\ldots, \vect{R}_N^T) = (x_1,p_1,\ldots,x_N,p_N),
\end{equation}
where $N$ stands for the number of vertices in the graph $N=|L|$. In this work, we will set $\hbar = 1$, and rescale the position and momentum operators such that they become dimensionless. Accordingly, the components of the vector $\vect{R}^T$ obey the usual canonical commutation relations
\begin{equation}
\label{cr_bosons}
  [\vect{R}_i, \vect{R}_j^T] = -\ii \delta_{ij} J,
  \end{equation}
where $i,j\in L$, $\delta_{ij}$ is the Kronecker delta, and we have defined $J=-\ii\sigma^y$ in terms of a Pauli matrix.

The spin degrees of freedom are represented by a set of dimensionless operators $\vect{S}_i=(S^1_i,\ldots,S^{m}_i)$ that form a Lie algebra. For concreteness, we  start by assuming that they form a representation of $\mathfrak{su}(2)$ with commutation relations
\begin{equation}
\label{cr_spins}
  [S_i^\alpha, S_j^\beta] = \ii \delta_{ij} \sum_{\gamma} f^{\alpha\beta\gamma} S_i^\gamma ,  
\end{equation}
where $\alpha,\beta,\gamma \in\{1,2,3\}$ label the different spin components, and $f^{\alpha\beta\gamma} = 2\epsilon^{\alpha\beta\gamma}$ is defined in terms of the completely antisymmetric   Levi-Civita symbol $\epsilon^{\alpha\beta\gamma}$. However, let us remark that the LRB derived below also applies to more general Lie algebras, as far as the structure constants $f^{\alpha\beta\gamma}$ are completely antisymmetric (see below).

\begin{figure}
\centering
\includegraphics[width=.55\columnwidth]{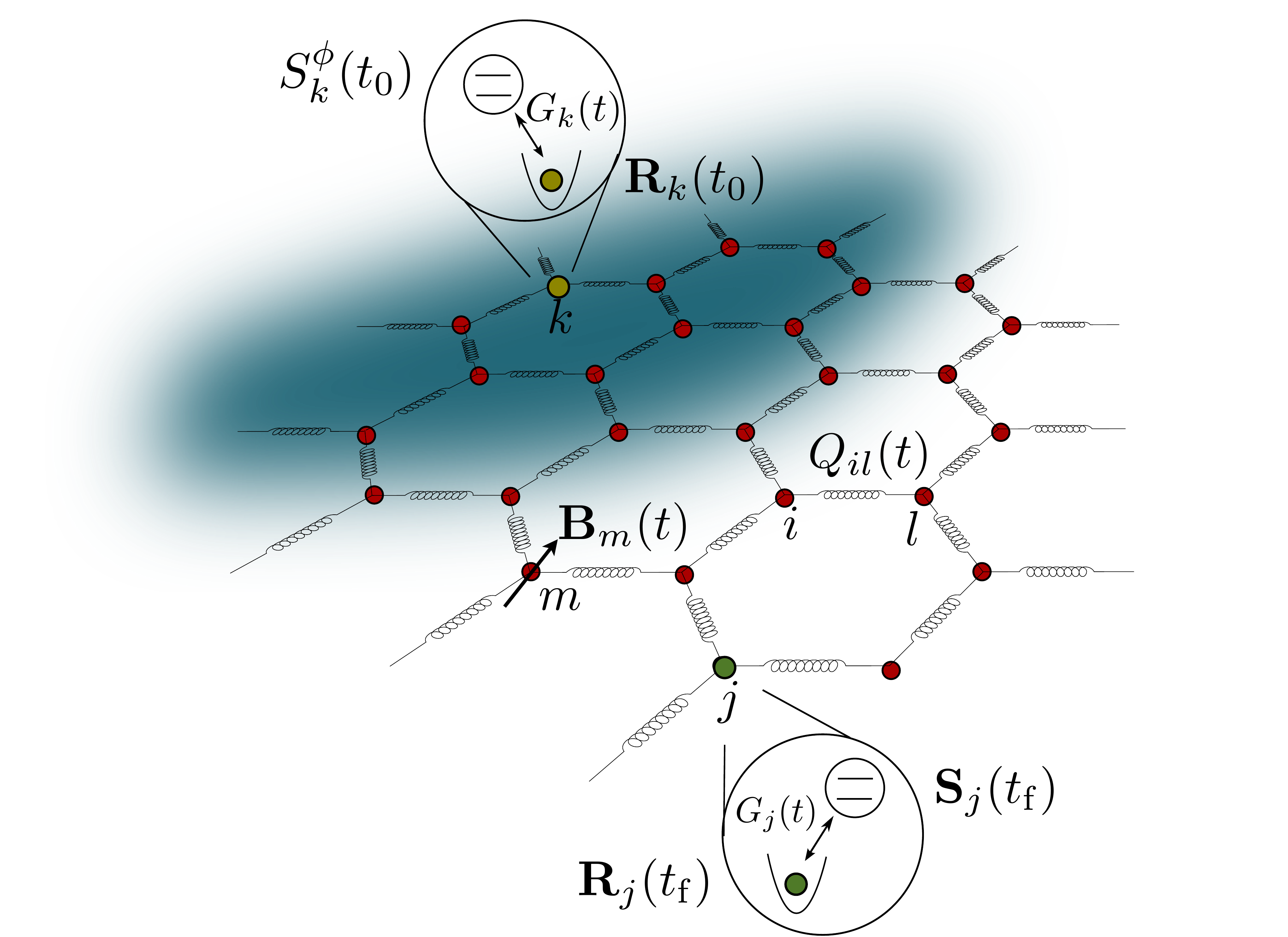}
\caption{ {\bf Scheme of the spin-boson lattice model: } Scheme of a graph corresponding to a honeycomb lattice with vertices $i,j,k,l\in L$  represented by red, green and yellow dots. In the two insets, we depict the spin $\vect{S}_j$ and bosonic $\vect{R}_j$ physical degrees of freedom by discrete levels and a quadratic well, respectively. The links of the lattice correspond to the edges of the graph $E$, and are represented as springs leading to the coupling $Q_{ij}(t)$ between distant oscillators. We also represent the on-site magnetic-field $\vect{B}_j(t)$ under which the spin precess, and the spin-boson coupling $G_{j}(t)$. The  blue region mimics the propagation of a spin perturbation at $S_k^{\phi}(t_0)$, until it reaches a distant spin $\vect{S}_j(t_{\rm f})$. }
\label{SBLM_scheme}
\end{figure}

The dynamics of this spin-boson lattice, and thus the LRB, will depend on a  particular choice for the Hamiltonian. Motivated by its applicability in different physical contexts (e.g. condensed matter), we will consider that bosons at distant lattice sites are coupled by a matrix $Q_{ij}(t)\in \mathcal{M}_{2\times2}(\mathbb{R})$ whose elements have the units of frequency (i.e. recall that $\hbar=1$), and fulfil $Q_{ij}(t)=Q_{ji}(t)\neq0$ whenever the two vertices $i,j\in L$ are connected through an edge of the graph.
The spins precess under a general magnetic field $\vect{B}^T_i(t)=(B_i^1(t),B^2_i(t),B^3_i(t))$, which might be time-dependent $B^{\alpha}_i(t)\in\mathbb{R}$, and also has the units of frequency.  Finally,  the coupling between spins and bosons is purely local (i.e. it takes place exclusively at the same vertex of the graph),  and is defined through the matrices $G_i(t) \in \mathcal{M}_{2\times 3}(\mathbb{R})$ that also have  the units of frequency.   Altogether, the Hamiltonian of the spin-boson lattice model is 
\begin{equation}
  H_{\textsf{SBL}} (t)= \frac{1}{2} \sum_{i,j} \vect{R}_i^T Q_{ij}(t) \vect{R}_j + \sum_i \vect{B}_i(t)^T\cdot \vect{S}_i+
  \sum_{i} \vect{R}_i^T \cdot G_{i}(t) \cdot \vect{S}_i.
  \label{full-H}
\end{equation}

Once the model has been defined, let us state the problem under study. Our objective is to understand how correlations are established between distant spins in the lattice, and to derive a bound on how fast this process can take place. Since distant spins do not interact directly, spin-spin interactions and thus spin correlations can only be mediated by the exchange of bosons, which form an array of coupled oscillators. This situation is common in physics, where bosons act as  carriers of the fundamental interactions between particles. Therefore, we believe that the results derived in this work are general enough to be of qualitative  interest to a wide variety of systems. Moreover, we will present a detailed description to argue that these results are also of quantitative importance for  crystals of trapped ions, a system in the field of quantum optics that offers a playground where our ideas can be tested experimentally.

Following the tradition of Lieb-Robinson bounds~\cite{hastings_les_houches}, we will study the so-called Lieb-Robinson commutators, whose expectation value  corresponds to a retarded spin-spin  correlation function for a particular state. Such a  commutator  relates a perturbation at  site $k\in L$ and instant $t_0=0$, with an observable at a distant site $j\in L$ and  $t>t_0$, namely
\begin{equation}
  \vect{Z}_{jk}(t) := [\mathbf{S}_j(t), S^\phi_k(0)],\label{correlators}
\end{equation}
where $\phi\in\{1,2,3\}$. Since the spins are coupled to the bosons, we shall also have to consider the correlations between spin and bosonic operators, which are related to the following Lieb-Robinson commutator
\begin{equation}
\label{correlators_spin_boson}
  \vect{C}_{jk}(t) := [\mathbf{R}_j(t), S^\phi_k(0)].
\end{equation}
The objective of this work is to derive a bound for the norm of the  Lieb-Robinson spin commutator 
\begin{equation}
\norm{\vect{Z}_{jk}(t)}\leq \xi(d_{jk},v_{\rm LR}t),
\label{LR_bound2}
\end{equation}
where $\xi(d_{jk},v_{\rm LR}t)$ is a certain function that depends on the distance $d_{jk}$ between the two lattice sites, and that also depends on time. Since for any state of the system, the retarded spin correlation function fulfils  
\begin{equation}
|C_{S_j^{\alpha},S_k^{\phi}}(t)|=|\big\langle [S^{\alpha}_j(t), S^\phi_k(0)]\big\rangle |\leq\norm{\vect{Z}_{jk}(t)}_{\infty}:={\rm max}_{\rm \beta}\norm{{Z}^\beta_{jk}(t)}, 
\label{correlator}
\end{equation}
we can also interpret that the function $f(d_{jk},v_{\rm LR}t)$ in the LRB~\eqref{LR_bound2} contains information on how fast  the spin correlations are established as the perturbation travels with a certain speed $v_{\rm LR}$ across the distance $d_{jk}$.

\subsection{Differential equations for the Lieb-Robinson commutators}

In order to study the commutators in Eq.~(\ref{correlators}), we will work with the Heisenberg picture. In this picture,  the time evolution of any operator $A$ is given by the following differential equation
$ \frac{\rm d}{{\rm d}t} A(t) =-\ii [A(t), H_{\textsf{SBL}}(t)]$.
Using the commutations relations in Eqs.~\eqref{cr_bosons} and~\eqref{cr_spins}, we arrive at the following system of ordinary differential equations (ODEs) for the boson and spin operators
\begin{align}
  \dot{\vect{R}}_j & = - \sum_{k} J \cdot Q_{jk}(t) \cdot \vect{R}_k - J \cdot G_j(t) \cdot \vect{S}_{j},
  \label{interaction-R}\\
 \dot{\vect{S}}_j & = \ii K_{j}(t) \cdot \mathbf{S}_j,
 \label{interaction-S}
\end{align}
where $K_j(t)$ is a matrix of operators  depending on the representation for the spins. For our particular choice, it can be written as
\begin{equation}
\label{precession}
K_j^{\alpha\beta}(t)=-\ii\sum_\gamma\big(\vect{B}_j(t)+\vect{R}_j^T \cdot G_j(t)\big)^\gamma f^{\alpha\gamma\beta},
\end{equation}
which can be easily shown to be Hermitian $K_j(t)=K_j^{\dagger}(t)$. 

\vspace{0.5ex}
{\it Evolution of the bosons.--} The first equation (\ref{interaction-R}) describes a set of coupled harmonic oscillators, where the spins act as a ``source'' term for the bosonic operators. This equation can be formally integrated (see Sect.~\ref{sec:lr-harmonic}) as
\begin{equation}
  \mathbf{R}_j(t) = \sum_kW_{jk} (t,0)\mathbf{R}_k(0) -
  \int_0^t \mathrm{d}\tau \sum_kW_{jk}(t,\tau) \cdot J \cdot G_k(\tau) \cdot \vect{S}_k(\tau),
  \label{eq-source}
\end{equation}
where we have used the propagator for the free bosons $W(t_1,t_2)$.
 According to the notation  above, we may regard $W(t_1,t_2)$ as an $N\times N$ block matrix, such that each of the blocks is a $2\times2$ matrix  $W_{jk}(t_1,t_2)$ that couples the oscillators at sites $j$ and $k$.

Following our results described in Sect.~\ref{sec:lr-harmonic}, which constitute an improvement with respect to the work of M. Cramer \textit{et al.} \cite{cramer_lrb_bosons}, the  norm of these propagators can be  bounded by
\begin{equation}
  \norm{W_{jk}(t,t_0)}\leq 
  e^{\nu_{\rm LR} |t - t_0|} \alpha \frac{ e^{-\mu d_{jk}}}{[1+d_{jk}]^\eta}
  =: \alpha e^{\nu_{\rm LR} |t - t_0|} f(d_{jk}),
\label{eq:harmonic-bound2}
 \end{equation}
where $v_{\rm LR}$ is the so-called Lieb-Robinson speed, which determines the maximum rate of propagation of bosonic perturbations in  the lattice. Note that this expression includes a more than exponential suppression for $\mu>0$ and a long-distance attenuation of $\norm{W_{jk}(t,t_0)}$ for $\mu=0$ due to the possibly long-range interactions in the oscillator couplings $Q_{jk}(t)$, a situation that will become evident when discussing  the trapped-ion realisation.

The only property that we will use is the fact that the spatial modulation $f(d)$ can be summed in the following way \begin{equation}
  a_0 := \max_{ik} \left[f(d_{ik})^{-1} \sum_j f(d_{ij}) f(d_{jk})\right] < +\infty,
  \label{definition_a_0}
\end{equation}
which can be interpreted as a bound on the convolution function and is used in different proofs of LRBs~\cite{cramer_lrb_bosons,nachtergaele}. For instance, if $\mu=0$ and one selects $d_{ij}$ as the graph distance (i.e. number of edges forming the shortest path connecting the two vertices $i,j\in L$), it is possible to estimate $\tilde{a}_0=\alpha c_{\rm D}2^{\eta+1}\zeta(1-D+\eta)$, where  $\zeta(s)$ is the Riemann zeta function, and $c_{\rm D}$ is a constant that depends on the particular graph of dimension $D$~\cite{cramer_lrb_bosons}. Such a constant can be determined by bounding the maximum number of vertices ${\rm sup}_{i}|S_{r}(i)|\leq c_Dr^{D-1}$ in a sphere of radius $r\in\mathbb{N}$, namely $S_r(i)=\{j\in L: d_{ij}=r\}$. Let us  remark, however, that the LRB can be made tighter for the type of lattices realised by the ion crystals, where the Euclidean distance arises naturally, and  will allow us to substitute $\tilde{a}_0\to a_0<\tilde{a}_0$.

\vspace{0.5ex}
{\it Evolution of the spins.--} The second equation~(\ref{interaction-S}) describes the precession of the spins under the operator $K_j$ which, in addition to the effects of the external magnetic field,   includes also the feedback from the bosonic subsystem. An important property  used below is that the hermiticity of $K_j$ implies that this operator can be regarded as the generator of unitary rotations $O_j(t)$, namely
\begin{equation}
  \label{spin-rotation}
  \frac{{\rm d}}{{\rm d}t} O_j(t) = \ii K_j(t) O_j(t),\quad O_j^{-1}(t)=O_j^{\dagger}(t).
\end{equation}

Let us note that this crucial property relies on the particular $\mathfrak{su}(2)$ commutation relations. In a more general case, where the matrices $\vect{S}_i$ have arbitrary dimension $d_i$, we may regard these matrices as acting on a subspace of a larger Hilbert space, $\mathbb{C}^{2^m},\, m \geq \lceil {\log d \over \log 2} \rceil$. In this even-sized space, we can find a set of Hermitian generators $S^{\alpha}_i\in \mathcal{M}_{2^m \times 2^m}(\mathbb{C}),$ that form a complete basis for the observable. The commutator of any two generators will depend, once more, on an antisymmetric tensor (see Ref.~\cite{different_spins}), which results from the composition of Levi-Civita symbols. Thanks to this fact, we can still prove that the operator $K_j(t)$ is the product of Hermitian operators ($x_n,\,p_n$) and a set of Hermitian matrices, obtaining, once more, that $O_j(t)$ is unitary.

\vspace{0.5ex}
{\it Evolution of the Lieb-Robinson commutators.--} Since we are actually interested in the commutators in Eqs.~\eqref{correlators}-\eqref{correlators_spin_boson}, we will write down their corresponding time-evolution equations. Note that $\vect{C}_{jk}(t)$ is not zero since the source terms in Eq.~(\ref{eq-source}) introduce  feedback of the spins into the oscillators, which  thus become correlated as the time evolves. Moreover, the opposite effect happens through the  $K_j(t)$, complicating the solution of the differential equations
\begin{align}
  \dot{\vect{C}}_{jk} &= - \sum_{l} J \cdot Q_{jl}(t)\cdot \vect{C}_{lk} - J \cdot G_j(t) \cdot \vect{Z}_{jk}\label{cs}\\
  \dot{\vect{Z}}_{jk} &= \ii \sum_n C_{jk}^n A^n_j(t)\cdot \vect{S}_j +\ii K_j(t)\cdot \vect{Z}_{jk},\label{zs}
\end{align}
where we have introduced  the index $n\in\{x,p\}$ to label the position/momentum spin-boson couplings, and the matrices $A^n_j(t)\in\mathcal{M}_{3\times3}(\mathbb{C})$, which have the following expression $A^n_j(t)=-\ii\sum_{\gamma} G_j^{n\gamma}(t)f^{\alpha\gamma\beta}$.

It is clear that the last term in Eq.~\eqref{zs}, namely $\ii K_j(t)\vect{Z}_{jk}$, cannot be  responsible for the propagation of spin correlations, as it is just a local  evolution of the spins that would be present even in the case of uncoupled oscillators. Fortunately, we have already shown that this local term  can be regarded as the generator of a unitary rotation~(\ref{spin-rotation}). Hence, we may eliminate this term by defining  a new set of spin-spin commutators $\vect{D}_{jk} := O^{-1}_j \vect{Z}_{jk}$, which share the norm with the original ones $\norm{\vect{D}_{jk}}=\norm{\vect{Z}_{jk}}$. The time-evolution for these spin commutators becomes
\begin{equation}
\label{ds}
  \dot{\vect{D}}_{jk} = \ii O_j^{-1}(t) \cdot \left [\sum_n C_{jk}^n \, A^n_j(t) \right ] \cdot \vect{S}_j,
\end{equation}
which is simple enough such that we can  derive the desired Lieb-Robinson bound for the spin-boson lattice model.

\subsection{ Lieb-Robinson bounds for the spin-boson lattice model}
 In analogy with the time-evolution of the bosonic operators~\eqref{eq-source}, the system of differential equations for the spin-boson commutator~\eqref{cs} can be formally integrated. Using the initial condition $\vect{C}_{jk}(0)={\bf 0}$, which assumes that the spins and bosons are initially uncorrelated, we arrive at 
\begin{equation}
  \vect{C}_{jk}(t) = -\int_0^t  \dd\tau \sum_{l}W_{jl}(t,\tau) \cdot J \cdot G_l(\tau) \cdot O_l(\tau)
  \cdot \mathbf{D}_{lk}(\tau).
\end{equation}
Upon substitution of this result in the system of ODEs for the spin-spin commutators~\eqref{ds}, we find
\begin{equation}
 \dot{\mathbf{D}}_{jk}(t) =\ii\sum_{l,n}  O_j^{-1}(t) \cdot
\int_0^t \dd\tau  \left [W_{jl}(t,\tau) \cdot J \cdot G_{l}(\tau) \cdot O_l(\tau) \cdot \mathbf{D}_{lk}(\tau) 
   \right ]^n \cdot A^n_j(t) \cdot \mathbf{S}_j(t).
\end{equation}
Integrating this equation leads to a Dyson-type recurrence that only contains spin operators
\begin{equation}
 {\mathbf{D}}_{jk}(t) = {\mathbf{D}}_{jk}(0) +\ii  \int_0^t\dd\tau_1\int_0^{\tau_1}\dd\tau_2 \sum_{l,n}O^{-1}_j(\tau_1) \cdot
 \left [W_{jl}(\tau_1,\tau_2) \cdot J \cdot G_{l}(\tau_2) \cdot
   O_l(\tau_2) \cdot \mathbf{D}_{lk}(\tau_2) 
 \right ]^n \cdot A^n_j(\tau_1) \cdot \mathbf{S}_j(\tau_1) .
\end{equation}
We can now upper-bound the norm of the Lieb-Robinson commutator $\vect{D}_{jk}(t)$ by using two properties of the operator norm, namely $\norm{A+B}\leq \norm{A}+\norm{B}$, and $\norm{AB}\leq \norm{A}\norm{B}$. Additionally, we will exploit the fact that $O_j$ is a unitary operator, $\norm{O_j(t)}=1$, and use the bound for the norm of the free bosonic propagator~\eqref{eq:harmonic-bound2}. After introducing the upper bounds $g=\max_{t,j}\norm{G_j(t)}=\max_{t,j,n}\norm{A^n_j(t)}$, and $S=\max_{t,j}\norm{\vect{S}_j(t)}$, we obtain
\begin{equation}
 \norm{{\mathbf{D}}_{jk}(t)} \le \norm{{\mathbf{D}}_{jk}(0)} +\chi\sum_{l} \int_0^t \dd\tau_1
 \int_0^{\tau_1}\dd\tau_2 \ee^{v_{\rm LR}(\tau_1-\tau_2)} f(d_{jl}) \norm{\mathbf{D}_{lk}(\tau_2)} .
\end{equation}
where we have introduced the constant  $ \chi=2 g^2  S \alpha$. We  now interchange the integration order, 
noting that the initial integration limits are $0 \leq \tau_2 \leq \tau_1$ and $0 \leq \tau_1 \leq t$, and defining the same integration region as $\tau_2 \leq \tau_1 \leq t$  and $0 \leq \tau_2 \leq t.$ Hence, 
\begin{align}
 \norm{{\mathbf{D}}_{jk}(t)} \le \norm{{\mathbf{D}}_{jk}(0)} +\chi\sum_{l} \int_0^t \dd\tau_2
 \int_{\tau_2}^{t}\dd\tau_1 \ee^{v_{\rm LR}(\tau_1-\tau_2)} f(d_{jl}) \norm{\mathbf{D}_{lk}(\tau_2)} .
\end{align}
Integrating the exponential term, we now find
\begin{align}
 \norm{{\mathbf{D}}_{jk}(t)} \le \norm{{\mathbf{D}}_{jk}(0)}+ \frac{\chi}{v_{\rm LR}}   \sum_{l} \int_0^t \dd\tau_2
\ee^{v_{\rm LR}(t-\tau_2)} f(d_{jl}) \norm{\mathbf{D}_{lk}(\tau_2)}.
\end{align}
Let us now iterate this recurrent expression, which leads us to
\begin{equation}
\begin{split}
 \norm{{\mathbf{D}}_{jk}(t)} \le \norm{{\mathbf{D}}_{jk}(0)}+ &
\frac{\chi}{v_{\rm LR}}   \sum_{l} \int_0^t \dd\tau_2 e^{v_{\rm LR} (t-\tau_2)} f(d_{jl}) \norm{\mathbf{D}_{lk}(0)}+
\\ +&\left(\frac{\chi}{v_{\rm LR}}\right)^2 \sum_{l,l'} \int_0^t \dd\tau_2\int_0^{\tau_2}\dd\tau_3 e^{v_{\rm LR} (t-\tau_3)} f(d_{jl}) f(d_{ll'}) \norm{\mathbf{D}_{l'k}(\tau_3)}. \\
\end{split}
\end{equation}
We now use the relation $\sum_{l}f(d_{jl}) f(d_{ll'})\leq a_0 f(d_{jl'})$ and we get
\begin{equation}
\begin{split}
 \norm{{\mathbf{D}}_{jk}(t)} \le \norm{{\mathbf{D}}_{jk}(0)}+ &
\left(\frac{\chi a_0}{v_{\rm LR}} \right)^{\phantom{2}}  \sum_{l} \int_0^t \dd\tau_2 e^{v_{\rm LR} (t-\tau_2)} \frac{1}{a_0}f(d_{jl}) \norm{\mathbf{D}_{lk}(0)}+\\ +&\left(\frac{\chi a_0}{v_{\rm LR}}\right)^2 \sum_{l'} \int_0^t \dd\tau_2\int_0^{\tau_2}\dd\tau_3 {e^{v_{\rm LR} (t-\tau_3)}} \frac{1}{a_0}f(d_{jl'}) \norm{\mathbf{D}_{l'k}(\tau_3)}. \\
\end{split}
\end{equation}

It is possible to iterate the above expression for $j\neq k$, using the equality $\norm{\mathbf{D}_{jk}(0)}=2\norm{\vect{S}_j}\norm{S_k^{\phi}}\delta_{jk}$, together with $\norm{\vect{S}_j}=\norm{\vect{S}_k}$ and $j\neq k$ which implies $\norm{{\mathbf{D}}_{jk}(0)} = 0$. As an example, the first three steps read
\begin{equation}
\begin{split}
	 &\norm{{\mathbf{D}}_{jk}(t)} \\
\le &\norm{{\mathbf{D}}_{jk}(0)}+ \left(\frac{\chi a_0}{v_{\rm LR}} \right)^{\phantom{2}}  \sum_{l} \int_0^t \dd\tau_2 \frac{e^{v_{\rm LR} (t-\tau_2)}f(d_{jl})}{a_0} \ \norm{\mathbf{D}_{lk}(0)}+\left(\frac{\chi a_0}{v_{\rm LR}}\right)^2 \sum_{l'} \int_0^t \dd\tau_2\int_0^{\tau_2}\dd\tau_3 \frac{e^{v_{\rm LR} (t-\tau_3)}f(d_{jl'})}{a_0}\ \norm{\mathbf{D}_{l'k}(\tau_3)}\\
\le & \: 2 \norm{\vect{S}_j}\norm{S^\phi_k}
 \left(\frac{\chi a_0}{v_{\rm LR}}\right)^{\phantom{2}} \frac{f(d_{jk})}{a_0}   \int_0^t  e^{v_{\rm LR} (t-\tau_2)}\dd\tau_2+\left(\frac{\chi a_0}{v_{\rm LR}}\right)^2 \sum_{l'} \int_0^t \dd\tau_2\int_0^{\tau_2}\dd\tau_3 \frac{e^{v_{\rm LR} (t-\tau_3)}f(d_{jl'})}{a_0}\ \norm{\mathbf{D}_{l'k}(0)} \\ &+\left(\frac{\chi a_0}{v_{\rm LR}}\right)^3 \sum_{l',l''} \int_0^t \dd\tau_2\int_0^{\tau_2}\dd\tau_3\int_0^{\tau_3}\dd\tau_4 \frac{e^{v_{\rm LR} (t-\tau_4)}f(d_{jl''})}{a_0}\ \norm{\mathbf{D}_{l''k}(\tau_4)} \\
\le & \: 2 \norm{\vect{S}_j}\norm{S^\phi_k}
 \left(\frac{\chi a_0}{v_{\rm LR}}\right)^{\phantom{2}} \frac{f(d_{jk})}{a_0}   \int_0^t  e^{v_{\rm LR} (t-\tau_2)}\dd\tau_2 + 2 \norm{\vect{S}_j}\norm{S^\phi_k}
 \left(\frac{\chi a_0}{v_{\rm LR}}\right)^2 \frac{f(d_{jk})}{a_0}  \int_0^t \dd\tau_2\int_0^{\tau_2}\dd\tau_3 e^{v_{\rm LR} (t-\tau_3)}\\ &+\left(\frac{\chi a_0}{v_{\rm LR}}\right)^3 \sum_{l''} \int_0^t \dd\tau_2\int_0^{\tau_2}\dd\tau_3\int_0^{\tau_3}\dd\tau_4 \frac{e^{v_{\rm LR} (t-\tau_4)}f(d_{jl''})}{a_0}\ \norm{\mathbf{D}_{l''k}(\tau_4)} \\
\end{split}
\end{equation}
The total sum up to infinite order reads
\begin{align}
 \norm{\mathbf{D}_{jk}(t)} &\le \sum_{n=1}^\infty  2 \norm{\vect{S}_j}\norm{S^\phi_k}
 \left(\frac{\chi a_0}{v_{\rm LR}}\right)^n \frac{\ee^{-\mu d_{jk}}f(d_{jk})}{a_0}T_n(j,k),
\end{align} with
\begin{equation}
 T_n = \int_0^t \dd \tau_1 \int_0^{\tau_1} \dd \tau_2 \cdots \int_0^{\tau_{n-1}} \dd \tau_n \, e^{v_{\rm LR}( t-\tau_n)}\le e^{v_{\rm LR} t }\int_0^t \dd \tau_1 \int_0^{\tau_1} \dd \tau_2 \cdots \int_0^{\tau_{n-1}} \dd \tau_n 
\end{equation} 
By direct integration,  we find that $T_n \le  e^{v_{\rm LR} t} \frac{{t}^n}{n!}$ which results in 
\begin{align}
 \norm{\mathbf{D}_{jk}(t)} &\le \sum_{n=1}^\infty  2 \norm{\vect{S}_j}\norm{S^\phi_k}
 \left(\frac{\chi a_0 t}{v_{\rm LR}}\right)^n{1 \over n!} \frac{\ee^{v_{\rm LR} t}}{a_0}f(d_{jk}).
\end{align}

The previous series can be summed up to infinite order yielding the desired Lieb-Robinson bound~\eqref{LR_bound2} for the spin-boson lattice model
\begin{equation}
\norm{\mathbf{Z}_{j,k}(t)}=\norm{\mathbf{D}_{j,k}(t)} \le 2 \norm{\vect{S}_j}\norm{S^\phi_k} \frac{e^{v_{\rm LR} t}f(d_{jk})}{a_0}\left(\ee^{\frac{\chi a_0}{v_{\rm LR}}t}-1\right), \hspace{2ex} \chi=2 g^2 S \alpha.
\label{LRB_sb}
\end{equation}
Interestingly enough, we find that the LRB for this composite system contains two contributions. On the one hand, the first exponential gives the maximum propagation speed for the bosons. Since the spin correlations are built by the exchange of bosons, it is natural that the speed of propagation of spin perturbations has an upper bound given by the speed of propagation of the carriers. On the other hand, the second exponential determines the efficiency with which distant spins can excite and reabsorb a propagating boson. Accordingly, this process should be proportional to $g$, the maximum spin-boson coupling strength. Moreover, if the bosons travel  much faster than the time-scale related to such a spin-boson coupling, an adiabatic-type argument tells us that the efficiency of excitation/reabsorption of bosons by distant spins should be reduced. Therefore, we can expect that, in addition, the process should also be proportional to $g/v_{\rm LR}$. These arguments based on a physical reasoning are confirmed by 
the rigorous LRB, as we have found that the argument of the second exponential is $\chi a_0/v_{\rm LR}\propto g^2/v_{\rm LR}$.


\section{Harmonic lattice Lieb-Robinson bounds}
\label{sec:lr-harmonic}

In this section of the Supplemental Material, we present an improvement on the LRB for free bosonic lattice models found by M. Cramer {\it et al.}~\cite{cramer_lrb_bosons}. We also show how this bound enters in the full spin-boson lattice model.

\vspace{0.5ex}
{\it Free oscillators.--} Let us rewrite the bosonic part of the full spin-boson Hamiltonian~\eqref{full-H} by separating the diagonal and off-diagonal terms
\begin{align}
 H_{\rm b} = \frac{1}{2}\sum_i \omega_i(t)\left(  p_i^2+ x_i^2\right) + \frac{1}{2}\sum_{i\neq j}  \vect{R}^T_i Q_{ij}(t) \vect{R}_j,\end{align}
where $\omega_i(t)=Q_{ii}(t)$, and  we impose that the   coupling matrices $Q_{ij}(t)$ between different lattice sites $i\neq j$ decay  with the distance. Moreover, we also assume that the norm of such matrices can be upper bounded by
\begin{align}
\label{bound_Q}
 \norm{Q_{ij}(t)}\le \frac{\kappa \ee^{-\mu d_{ij}}}{(1+d_{ij})^\eta},
\end{align}
where we use the exponent $\eta\in\mathbb{Z}$, and a constant $\kappa\in\mathbb{R}$ with the units of frequency. For $\mu>0$, the couplings are more than exponentially suppressed; for $\mu=0$, we face an algebraic decay.

We can now get rid of the diagonal terms by changing into a frame in which the phase space coordinates rotate with angular speed $\omega_j(t)$
\begin{equation}
  \vect{R}_j(t) =
  U_j(t) \tilde{\vect{R}}_j (t),\hspace{2ex} U_j(t)=\ee^{-\int_{t_0}^t \dd\tau \omega_j(\tau)J}
\end{equation}
In this frame, the system of ODEs for the free bosonic operators~\eqref{interaction-R} only includes the off-diagonal couplings 
\begin{align}
  \frac{\dd {\bf{\tilde{R}}}_j}{\dd t} = - \sum_{k\neq j} U_j^{-1}(t)\cdot J \cdot Q_{jk}(t)\cdot U_k(t) \cdot \vect{\tilde{R}}_k,
\end{align} 
which can be alternatively written in terms of the free bosonic propagator $\vect{\tilde{R}}(t) = \tilde{W}(t,t_0)\vect{\tilde{R}}(t_0)$ in the rotated frame, such that the full propagator corresponds to  ${W}_{jk}(t,t_0)=U_{j}(t,t_0)\tilde{W}_{jk}(t,t_0).$ This propagator satisfies the Dyson series
\begin{equation}
  \tilde{W}_{jk}(t,t_0) = \delta_{jk} \openone -
  \sum_{l \neq j} \int_{t_0}^t \dd\tau U_j^{-1}(t)\cdot J \cdot Q_{jl}(t)\cdot U_l(t)  \cdot \tilde{W}_{l,k}(\tau,t_0) ,
\end{equation}
where we used $\tilde{W}(t_0,t_0)=\openone$ for all $t_0$. Let us calculate the norm of this operator, and use the identities $\norm{A+B}\leq \norm{A}+\norm{B}$, and $\norm{AB}\leq \norm{A}\norm{B}$. Moreover, since $U_j(t),$ and $J$ are  unitary operators, it follows that
\begin{equation}
  \norm{W_{jk}(t,t_0)} \leq \delta_{jk} +
  \sum_{l \neq j} \int_{t_0}^t \dd\tau \norm{Q_{jl}(t)} \norm{W_{lk}(\tau,t_0)}.
\end{equation}
By using the bound on the off-diagonal couplings~\eqref{bound_Q}, we find the expression
\begin{equation}
  \norm{W_{jk}(t,t_0)} \leq \delta_{jk} +
  \sum_{l \neq j}\frac{\kappa\ee^{-\mu d_{jl}}}{(1+d_{jl})^{\eta}} \int_{t_0}^t \dd\tau\norm{W_{lk}(\tau,t_0)},
\end{equation}
which can be iterated once  to obtain
\begin{equation}
  \norm{W_{jk}(t,t_0)} \leq \delta_{jk} +
  \sum_{l \neq j}\frac{\kappa \ee^{-\mu d_{jl}}}{(1+d_{jl})^{\eta}} \int_{t_0}^t \dd\tau_1\delta_{lk}+\sum_{l\neq j}\sum_{l' \neq l}\frac{\kappa\ee^{-\mu d_{jl}}}{(1+d_{jl})^{\eta}}\frac{\kappa\ee^{-\mu d_{ll'}}}{(1+d_{ll'})^{\eta}}  \int_{t_0}^t \dd\tau_1\int_{t_0}^{\tau_1} \dd\tau_2\norm{W_{l'k}(\tau_2,t_0)}.
\end{equation}
In analogy to Eq.~\eqref{definition_a_0}, by introducing the geometric factor 
\begin{equation}
\tilde{a}_0 = \max_{jl'}\left\{e^{\mu d_{jl'}} (1+d_{jl'})^\eta
\sum_{l\neq j}e^{-\mu d_{jl}} (1+d_{jl
})^{-\eta}e^{-\mu d_{ll'}} (1+d_{ll'
})^{-\eta}\right\},
\end{equation}
we find directly that
\begin{equation}
  \norm{W_{jk}(t,t_0)} \leq \delta_{jk} +
  (\kappa \tilde{a}_0)\frac{\ee^{-\mu d_{jk}}}{\tilde{a}_0(1+d_{jk})^{\eta}} \int_{t_0}^t \dd\tau_1+(\kappa \tilde{a}_0)^2\sum_{l' \neq l}\frac{\ee^{-\mu d_{jl'}}}{\tilde{a}_0(1+d_{jl'})^{\eta}}  \int_{t_0}^t \dd\tau_1\int_{t_0}^{\tau_1} \dd\tau_2\norm{W_{l'k}(\tau_2,t_0)}.
\end{equation}
 By iterating this recursion to infinite order, we can now see that the free boson propagator can be expressed as 
 \begin{equation}
  \norm{W_{jk}(t,t_0)} \leq \delta_{jk}+\frac{\ee^{-\mu d_{jk}}}{\tilde{a}_0(1+d_{jk})^{\eta}}\sum_{n=1}^{\infty}(\kappa \tilde{a}_0)^n\tilde{T}_n,\hspace{2ex}
 \tilde{T}_n =\int_0^t \dd \tau_1 \int_0^{\tau_1} \dd \tau_2 \cdots \int_0^{\tau_{n-1}} \dd \tau_n.
 \end{equation}
 In this case, we find exactly that $\tilde{T}_n=t^n/n!$, such that the series can be summed to infinite order yielding 
 \begin{equation}
 \label{w_bound}
   \norm{W_{jk}(t,t_0)} \leq \delta_{jk} +
\frac{\ee^{\kappa \tilde{a}_0t-\mu d_{jk}}}{\tilde{a}_0(1+d_{jk})^{\eta}}\leq \frac{(1+\tilde{a}_0)}{\tilde{a}_0}\frac{\ee^{\kappa \tilde{a}_0t -\mu d_{jk}}}{(1+d_{jk})^{\eta}}.
 \end{equation}
This is precisely the LRB for the free bosonic lattice model that has been used in Eq.~\eqref{eq:harmonic-bound2} (note that the propagator can be directly related to the Lieb-Robinson retarded commutators of position-momentum operators~\cite{cramer_lrb_bosons}). We observe that the bound inherits the decay structure of the couplings. Regarding  algebraically decaying couplings ($\mu = 0$), this bound presents certain improvements with respect to the work of M. Cramer {\it et al.}~\cite{cramer_lrb_bosons}. The first and most important one is that the Lieb-Robinson speed $v_{\rm LR}=\kappa \tilde{a}_0$  only depends on the bound of the off-diagonal couplings~\eqref{bound_Q} thorough $\kappa$. As expected from a physical reasoning, the maximum speed with which the bosonic correlations are build up cannot increase  with the on-site frequencies, but must be rather limited by the off-diagonal couplings between distant oscillators. The second reason for the improvement  is that our bound applies to more general 
bosonic lattice models, where the coupling matrix $Q_{ij}(t)$ might depend on time, and  have all types of couplings, namely  position-position, position-momentum, and momentum-momentum couplings.

\vspace{0.5ex}
{\it Oscillators with source.--} In this part of the Supplemental Material, we  describe how the system of ODEs for the oscillators are solved according to Eq.~\eqref{eq-source} when they contain an additional source term $ \vect{F}(t)$, namely
\begin{equation}
  \dot{ \vect{R}}_j = - \sum_{k}J \cdot Q_{jk}(t) \cdot \vect{R}_k + \vect{F}(t).
\end{equation}
We can solve it with the following ansatz $
  \vect{R}(t) = W(t,t_0) [\vect{R}(t_0) + \vect{X}(t)],
$
where $W(t,t_0)$ is the propagator of the free bosonic system (i.e. homogeneous system of ODEs) previously bounded in Eq.~\eqref{w_bound}.
It is important to remark that we need both the starting time and the final time in $W$ because $Q(t)$ is time dependent: we have lost translational invariance in time. When we introduce this ansatz into the previous equation and impose that $W$ is the propagator of the free bosonic system,
$
  \frac{d}{dt} W(t,t_0) = Q(t) W(t,t_0),
$
we obtain $W(t,t_0) \frac{d}{dt} \vect{X}(t) = \vect{F}(t)$.
This leads to the solution
\begin{equation}
  \vect{R}(t) = W(t,t_0)\vect{R}(t_0) + W(t,t_0) \int_{t_0}^t \dd\tau W(\tau,t_0)^{-1} \vect{F}(\tau) .
\end{equation}
Using the fact that the operators can be composed, i.e. $W(t,t_0)=W(t,\tau)W(\tau,t_0)$, which follows from the properties of the  solution of the homogeneous system of ODEs, we can simplify the previous expression
\begin{equation}
  \vect{R}_j(t) =\sum_k W_{jk}(t,t_0)\vect{R}_k(t_0) + \int_{t_0}^t \sum_{k}W_{jk}(t,\tau) \vect{F}_k(\tau) \dd\tau.
\end{equation}
Finally, by considering the particular spin-dependent source term, $\vect{F}_k(\tau)=-J\cdot G_k(\tau)\cdot \vect{S}_k(\tau)$, we get the desired formal solution used in Eq.~\eqref{eq-source} above.


\section{ Lieb-Robinson bounds for spin correlations in trapped-ion crystals}

In this section of the Supplemental Material, we provide a detailed description of the applicability of the Lieb-Robinson bound (LRB) for  spin-boson lattice models in a trapped-ion system.

\subsection{Spin-boson lattice models with crystals of trapped ions}

Paralleling our discussion in Sect.~\ref{sec:statement}, let us start this section by describing how the general spin-boson lattice Hamiltonian~\eqref{full-H} can be realised in  state-of-the-art experiments with trapped ions~\cite{haeffner_review2}. We consider a collection of $N$ atomic ions of mass $m$, and charge $e$, confined in either  {\it (i)} a linear Paul trap, {\it (ii)} a Penning trap, or {\it (iii)} a micro-fabricated surface trap (see~\cite{traps}). For low-enough temperatures, the ions crystallise  forming either {\it (i)} a one-dimensional chain, a  {\it (ii)} triangular lattice in the rotating laboratory frame,  or {\it (iii)} any desired two-dimensional lattice. The equilibrium positions  ${\bf r}_i^0$, where  $i\in\{1,\ldots,N\}$  labels the different ions, correspond to the physical realisation of the set of vertices $L$ of the graph $G$  of Sect.~\ref{sec:statement} (see Fig.~\ref{SBLM_scheme}). Besides,  the set of edges $E$ is determined by the all possible links for each  lattice. 
The physical degrees of freedom of each vertex $i\in L$ correspond to the small transverse vibrations  of the ions around the equilibrium positions (i.e. bosons), living in $\mathcal{L}^2(\mathbb{R})$, and to a pair of  internal levels of the atomic level structure (i.e. spins),  living in $\mathcal{H}_i=\mathbb{C}^2$.

The small transverse vibrations will be denoted as $\delta r_{i,{\rm t}}$. They correspond to {\it (i)} one of the two directions perpendicular to the axis of the linear Paul trap, or to {\it (ii)-(iii)}  the direction perpendicular to the crystal plane in the Penning or surface traps. For any of these configurations, the transverse vibrations  decouple from the remaining vibrations of the ion crystal, and can be thus described by the same harmonic Hamiltonian, namely
\begin{equation}
\label{tv_ham}
H_{\rm b}=\sum_i\left(\frac{p^2_{i,{\rm t}}}{2m}+\frac{m}{2}\omega_{\rm t}^2\delta r_{i,{\rm t}}^2\right)+\frac{1}{2}\sum_{i,j}\mathbb{V}_{ij}\delta r_{i,{\rm t}}\delta r_{j,{\rm t}}.
\end{equation}
Here, the couplings between distant ions are obtained by expanding the Coulomb potential to second order in the small transverse displacements, which leads to  $\mathbb{V}_{ij}=e_0^2/|\vect{r}_i^0-\vect{r}_j^0|^3$ for $i\neq j$, and $\mathbb{V}_{ii}=-\sum_{j\neq i}\mathbb{V}_{ij}$, where $e_0^2=e^2/4\pi\epsilon_0$. Note also that the origin of the trap frequency $\omega_{{\rm t}}$ shall depend on the particular trap (i.e. for {\it (i)}-{\it (iii)} $\omega_{{\rm t}}$ is proportional to the r.f. frequency, whereas for {\it (ii)} it is proportional to the d.c. potential). By rescaling the position and momentum operators, $x_i=\sqrt{m\omega_{{\rm t}}}\delta r_{i,{\rm t}}$ and $p_i=p_{i,{\rm t}}/\sqrt{m\omega_{{\rm t}}}$, we can define the bosonic operators $\vect{R}_i^{T}=(x_i,p_i)$ with the commutation relations in Eq.~\eqref{cr_bosons}. Moreover, the Hamiltonian for the transverse vibrations~\eqref{tv_ham} can be rewritten as
\begin{equation}
\label{Q-H}
H_{\rm b}=\frac{1}{2}\sum_{i,j}\vect{R}_i^T\cdot Q_{ij}\cdot \vect{R}_j,\hspace{2ex} Q_{ij}=\left(\begin{array}{cc}\omega_{\rm t}\delta_{ij}+\frac{\mathbb{V}_{ij}}{m\omega_{\rm t}} & 0 \\0 & \omega_{\rm t}\delta_{ij}\end{array}\right),
\end{equation}
which yields a transparent realisation of the free bosonic part~\eqref{full-H}. It is also worth pointing out that the trap frequencies could be modified dynamically in the experiment to study e.g. quenches, leading to a time-dependent $Q_{ij}(t)$  also captured by our LRB. 

Let us now turn into the spin degrees of freedom, which correspond to a pair of atomic levels $\{\ket{\uparrow_i},\ket{\downarrow_i}\}$ with a sufficiently long coherence time. For the sake of concreteness, we select two  states from the hyperfine ground-state manifold of a certain ion (e.g. $^{{9}}{\rm Be}^+$ or $ ^{25}{\rm Mg}^+$), although we emphasise that the LRB~\eqref{LRB_sb} would equally apply to optical or Zeeman spins (e.g. $ ^{40}{\rm Ca}^+$ or $ ^{88}{\rm Sr}^+$). By defining the Pauli matrices $\sigma_i^x=\ket{{\uparrow_i}}\bra{{\downarrow_i}}+{\rm H.c.}$, $\sigma_i^y=-\ii\ket{{\uparrow_i}}\bra{{\downarrow_i}}+{\rm H.c.}$, and $\sigma_i^z=\ket{{\uparrow_i}}\bra{{\uparrow_i}}-\ket{{\downarrow_i}}\bra{{\downarrow_i}}$, it follows directly that the desired commutation relations~\eqref{cr_spins} are fulfilled. We define $\omega_0$ as the transition frequency between the two atomic states, and use electromagnetic radiation (e.g. a Raman configuration with two co-propagating laser beams, or a single 
microwave in a traveling-wave configuration), such that its  frequency fulfils $\nu\approx \omega_0$, and $|\nu-\omega_0|\ll \omega_{0}$. Then, the light-matter interaction reads as follows
\begin{equation}
\label{spin_b_field}
H_{\rm s}=\sum_i\vect{B}_i^T(t)\cdot\boldsymbol{\sigma}_i,\hspace{3ex}B_i^x(t)=\Omega\cos(\nu t-\varphi),\hspace{1ex} B_i^y(t)= \Omega\sin(\nu t-\varphi), \hspace{1ex}  B_i^z(t)=\frac{\omega_0}{2}, 
\end{equation}
where  $\boldsymbol{\sigma}_i=({\sigma}_i^x,{\sigma}_i^y,{\sigma}_i^z)^{T}$. Here, $\Omega\in\mathbb{R}$ is the so-called Rabi frequency  of the transition~\cite{qo_book}, and $\varphi$ depends on the phases of the  electromagnetic wave and the atomic dipole element.
The above expression corresponds to the free spin part of Eq.~\eqref{full-H}.

The only missing ingredient of the target lattice Hamiltonian~\eqref{full-H} is the spin-boson coupling. This requires a light-matter interaction that couples the internal states of the ions to the transverse vibrations, which can be achieved via the so-called state-dependent  dipole forces. Such forces  are nowadays routinely used for quantum-information processing in the trapped-ion community. For hyperfine spins,  by employing the gradient of either a travelling wave in a Raman configuration with non-co-propagating lasers~\cite{laser_state_dep_forces}, or  an oscillating magnetic field in the near-field of a microwave source~\cite{microwave_state_dep_forces}, it is possible to obtain 
\begin{equation}
\label{sd_force}
H_{\rm sb}=\sum_{i}g {x}_i\sigma_{i}^{z}\sin(\tilde{\nu}t-\tilde{\varphi}).
\end{equation} In this expression, $g=\sqrt{2}\tilde{\Omega}\gamma$ is the coupling strength, which  shall be referred to as the force  (note, however, that this force has the units of frequency since the rescaled position operator is dimensionless and $\hbar=1$). In the expression of the force,  $\tilde{\Omega}$ is the crossed-beam ac-Stark shift that originates from the Raman laser beams, or an ac-Zeeman shift in the near-field of an oscillating microwave source. We have also defined the so-called Lamb-Dicke parameter $\gamma\ll 1$, which depends on the gradient of the modulation of the electric (magnetic) field of the laser (microwave) at the position of the ion, and the zero-point motion of the ions. In order to get such state-dependent force, we have to consider that $\tilde{\nu}\approx\omega_{\rm t}$, such that $|\tilde{\nu}-\omega_{\rm t}|\ll\omega_{\rm t}$ and $|\tilde{\Omega}|\ll\omega_{\rm t}$, in order to make the gradient of the light-matter interaction dominant, which leads to Eq.~\eqref{sd_force} as opposed to the case in Eq.~\eqref{spin_b_field}. We can rewrite the state-dependent forces using the notation in Eq.~\eqref{full-H} as follows
\begin{equation}
\label{G-H}
H_{\rm sb}=  \sum_{i} \vect{R}_i^T \cdot G_{i}(t) \cdot \boldsymbol{\sigma}_i,\hspace{2ex} G_{i}(t)=\left(\begin{array}{ccc}0 & 0 & g \sin(\tilde{\nu}t-\tilde{\varphi}) \\0 & 0 & 0\end{array}\right),
\end{equation}
which yields a transparent realisation of the  spin-boson coupling of Eq.~\eqref{full-H}. Accordingly, Eqs.~\eqref{Q-H},\eqref{spin_b_field}, and~\eqref{G-H} are the ingredients for the trapped-ion realisation of the spin-boson lattice model 
\begin{equation}
H_{\textsf{SBL}}(t)=H_{\rm b}+ H_{\rm s}+ H_{\rm sb}.
\label{sb_ions}
\end{equation} 
As the form of the trapped-ion Hamiltonian $H_{\textsf{SBL}}(t)$ coincides exactly with the general model in Eq.~\eqref{full-H}, we can use directly the LRB~\eqref{LRB_sb} derived in the previous sections. This will allow us to estimate the maximum speed at which spin correlations can build up in a trapped-ion experiment. Before closing this section, let us remark once more that  the spin-boson dynamics of all these  ion crystals in the different traps (i.e. linear Paul trap, Penning trap, or surface trap) is described by the same Hamiltonian. Therefore, the LRB may find a broad application in a variety of ion-trap setups. We should also point out that, although the state-dependent force~\eqref{sd_force} corresponds to the so-called $\sigma^z$-force, other configurations lead to state-dependent forces in the $\sigma^x$-, or $\sigma^y$-bases~\cite{laser_state_dep_forces}. Note that our LRB~\eqref{LRB_sb}  would apply equally to any of these cases.

\subsection{Lieb-Robinson bounds for trapped-ion crystals}

\subsubsection{Bounds for  non-perturbative spin-boson models}

In this section, we will apply the LRB~\eqref{LRB_sb} for the trapped-ion spin-boson lattice model. For reasons that will become clear in Sect.~\ref{sect_measurment}, we will have experimental access to the retarded spin correlation functions (see Eq.~\eqref{correlator}). Therefore, we need to evaluate the the following bounds for the supremum norms 
\begin{equation}
\norm{G_i(t)}_{\infty}\leq g, \hspace{2ex}\norm{\boldsymbol{\sigma}_i}_{\infty}\leq S=1,\hspace{2ex} \norm{Q_{ij}}_{\infty}\leq\frac{8\beta\omega_{\rm t}}{(1+d_{ij})^3}, \hspace{1ex}\forall i\neq j.
\label{bounds}
\end{equation}
 Here, we have used the  stiffness parameter~\cite{porras_ising}, which measures the ratio of the Coulomb repulsion to the trapping energy, $\beta=e_0^2/m\omega_{\rm t}^2 {d}_{\rm m}^3$, where ${d}_{\rm m}={\rm min}_{i,j}\{|\vect{r}_i^0-\vect{r}_j^0|\}$ is the minimum distance between two ions in the crystal. We note that $\beta\ll 1$ for the setups considered in this work, which corresponds to a tight transverse confinement.  In this expression, we use $d_{ij}$ as the Euclidean distance between two vertices $i,j\in L$ of a perfect Bravais lattice, which has unit primitive vectors, and shares the  geometry with the ion crystal (i.e. note that ion crystals are usually characterised by an inhomogeneous lattice spacing). According to the above quantities, the LRB~\eqref{LRB_sb} for the spin-boson lattice model in an ion crystal is given by
\begin{equation}
\norm{ [\vect{\sigma}_i(t), \sigma^{\phi}_j(0)] }_{\infty}\leq {2 \over a_0 (1 + d_{ij})^3} \ee^{8a_0 (\beta\omega_{\rm t})t} \bigg( e^{\alpha\left(\frac{g}{ \beta\omega_{\rm t}}\right)^2(\beta\omega_{\rm t})t} -1 \bigg), \hspace{2ex}\alpha=\frac{1}{4}\left(\frac{1+a_0}{a_0}\right),
\label{lrb_ions_spin_boson2}
\end{equation} 
where $\beta\omega_{{\rm t}}$ is the typical order of magnitude for the tunneling of vibrational excitations between neighbouring ions. Therefore, the LR speed for the bosons is related to this tunneling, which is in fact the underlying mechanism responsible for the spread of correlations in both the free bosonic system, and the spin-boson model. However,  if we do not allow for a time that is sufficiently long  such that bosons can be exchanged between  the spins (i.e. $t> (g^2/ \beta\omega_{\rm t})^{-1}$), no correlations can build up regardless of how fast  the vibrational excitations propagate (i.e. the  term between brackets makes the correlations negligible, see Fig.~\ref{fig_LRB_1d}\,{\bf (a)}).

Let us now evaluate evaluate the LRB by considering realistic parameters for the different ion crystals of interest. First of all, we need to obtain the constant $a_0$, which is defined through the following bound of the convolution
$\sum_{l}(1+d_{il})^{-3}(1+d_{lj})^{-3}\leq a_0(1+d_{ij})^{-3}, \,\forall\, i,j\in L$. Alternatively, we can define
\begin{equation}
a_0={\rm max}_{i,j}\left\{\sum_{l\in L}\frac{(1+d_{il})^{-3}(1+d_{lj})^{-3}}{(1+d_{ij})^{-3}}\right\},
\label{maximization}
\end{equation}
a maximisation problem that will be solved for the crystals of interest.


\begin{figure}
\centering
\includegraphics[width=.7\columnwidth]{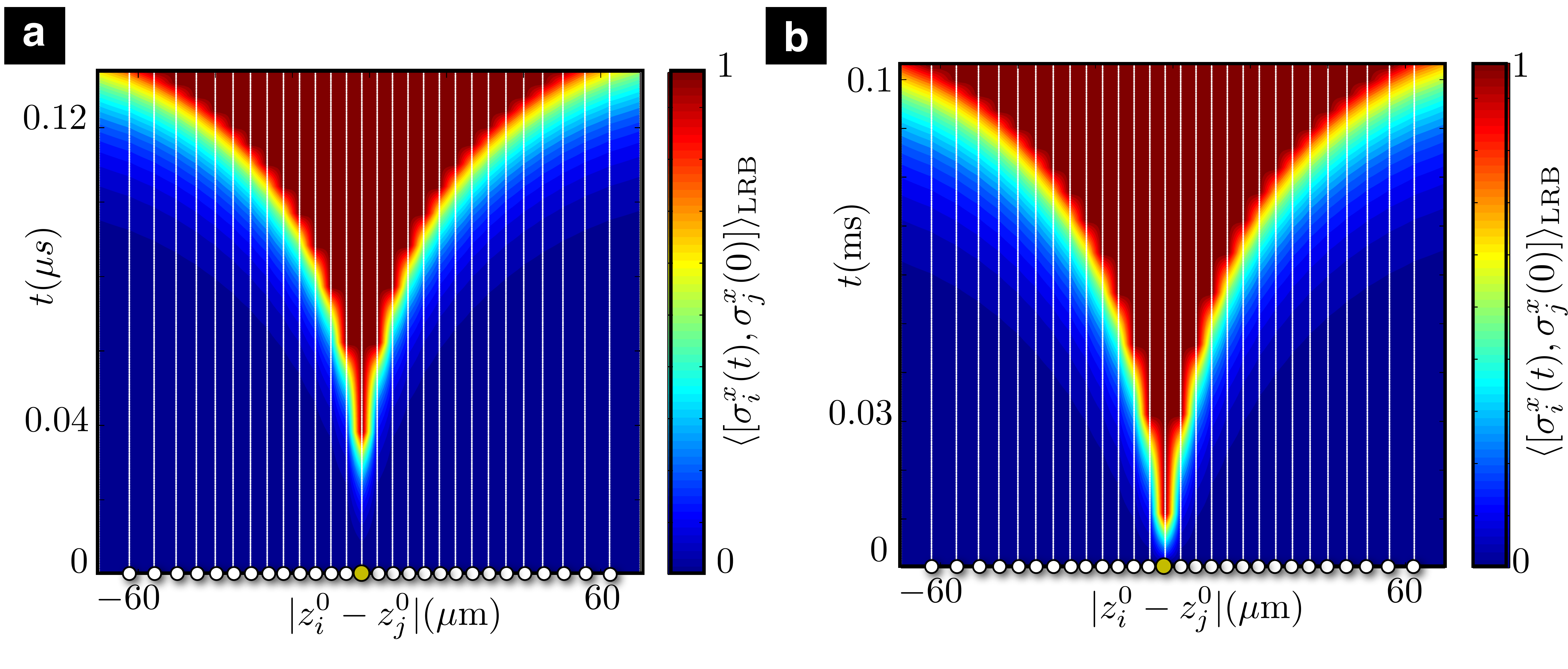}
\caption{ {\bf LRB for the spin correlations in an  ion chain:} {\bf (a)} Evaluation of the spin-boson LRB in Eq.~\eqref{lrb_ions_spin_boson2} for a linear  chain with $N=30$ $^{25}{\rm Mg}^+$ ions in a linear Paul trap (see the text for the specific parameters). The spin excitation initially localised at the middle of the chain, $j=N/2$, propagates towards the edges giving rise to a quasi-LR cone. We also observe that the cone requires a finite time to arise, which corresponds to the required time to create/annihilate bosons at distant sites. {\bf (b)} Evaluation of the spin LRB in Eq.~\eqref{lrb_ions_spins} for a linear  chain with $N=30$ $^{25}{\rm Mg}^+$ ions in a linear Paul trap (see the text for the specific parameters). We observe an analogous quasi-LR cone, where one must appreciate the very different time-scale of correlation spread as compared to the spin-boson LRB in Eq.~\eqref{lrb_ions_spin_boson2}  displayed in {\bf (a)}.}
\label{fig_LRB_1d}
\end{figure}

\vspace{0.5ex}
{\it (i) Ion chain in a linear Paul trap.--} In this case, the perfect Bravais lattice associated to the inhomogeneous ion chain is spanned by ${\bf a}_1={\bf e}_z$, such that $\tilde{{\bf r}}^0_i=i{\bf a}_1$, where $i\in\mathbb{Z}$. Hence, the Euclidean distance is simply $d_{ij}=|i-j|$, and we can maximise the above expression~\eqref{maximization} numerically to find that $a_0=2.9$. Let us note that this constant differs from the generic estimate~\cite{cramer_lrb_bosons} based on the graph distance $\tilde{a}_0=c_{1}2^{4}\zeta(3)= 38.5$ by an order of magnitude, a fact that  will make our LRB much tighter.

We now consider realistic parameters at reach of current ion-trap experiments. We consider $^{25}{\rm Mg}^+$ ions   in a linear Paul trap with frequencies $\omega_{\rm ax}/2\pi=0.25\,$MHz, and $\omega_{\rm t}/2\pi=5\,$MHz (see e.g.~\cite{schaetz}). This trap confines $N=30$ ions forming a linear chain of length $\ell\approx 140\,\mu$m, such that the minimum ion distance occurs at the centre of the trap $d_{\rm m}\approx 4\,\mu$m, and  the tunneling rate of vibrational excitations is $\beta\omega_{\rm t}/2\pi\approx 450\,$kHz. Finally, we need to estimate the value of the state-dependent force, $g=\sqrt{2}\tilde{\Omega}\gamma$. Considering that $\gamma\approx0.14$, and that $|\tilde{\Omega}|\ll\omega_{\rm t}$, it seems reasonable to consider that the force can be pushed towards $g/2\pi=100\,$kHz. In this regime, we find that the LRB~\eqref{lrb_ions_spin_boson2} corresponds to the  spin correlation spread displayed in Fig.~\ref{fig_LRB_1d}\,{\bf (a)}. Due to the long range  of the vibrational couplings, 
instead of a perfect Lieb-Robinson cone, we recover  a quasi-LR-cone. In any case, it is clear that there is a maximum propagation speed for spin correlations in such a spin-boson medium, and distant spins require a certain minimal time after which correlations can start building up. It is important to note that the timescale of correlation propagation of the LRB is in the $\mu$s range even for long chains of $N=30$, a timescale that is short enough such that other sources of noise (e.g. magnetic-field noise or heating) can be safely neglected. 

Although we have considered the particular case of $^{25}{\rm Mg}^+$ ion chains, we emphasise that similar experiments can be carried out with other ion species. In fact, linear chains  with up to $N=6$ ions of $^{40}{\rm Ca}^+$~\cite{lanyon}, and $N=3$~\cite{wunderlich} or  $N=16$~\cite{monroe} ions of $^{171}{\rm Yb}^+$ have already been used in experiments for digital and analog quantum simulations of transverse Ising models. These models arise from  spin-boson Hamiltonians  equivalent to Eq.~\eqref{sb_ions}, in a certain regime where the boson can be traced out (see Sect.~\ref{spin_spin}). Therefore,  in order to test the bound displayed in Fig.~\ref{fig_LRB_1d}\,{\bf (a)}, one would need to consider larger ion chains, and non-perturbative regimes where the ion crystal forms a spin-boson medium.

\vspace{0.5ex}
{\it (ii) Triangular lattice in a Penning or surface trap.--} In this case, the equivalent Bravais lattice is spanned by ${\bf a}_1={\bf e}_x$, and ${\bf a}_2={\bf e}_x/2+\sqrt{3}{\bf e}_y/2$, such that $\tilde{\bf r}^{0}_{\bf i}=i_1{\bf a}_1+i_2{\bf a}_2$ and ${\bf i}=(i_1,i_2)\in\mathbb{Z}\times\mathbb{Z}$. By  maximising~\eqref{maximization} with the Euclidean distance $d_{\bf i j}=|\tilde{\bf r}^{0}_{\bf i}-\tilde{\bf r}^{0}_{\bf j}|$, we find $a_0=8.5$. Once again, the estimate based on the graph distance would give give $\tilde{a}_0=c_{2}2^{4}\zeta(4)= 103.9$ overestimating the LR speed  by   one order of magnitude. Let us now specify the remaining parameters to evaluate the LRB. 

We  start by considering the experimental values for a $^9{\rm Be}^+$ crystal in a Penning trap~\cite{britton}, where  the transverse trap frequency is $\omega_{\rm t}/2\pi\sim0.8\,$MHz, and typical distances are $d_{\rm m}\sim20\,\mu$m. For these parameters, we can estimate that the tunneling of transverse vibrational excitations is on the order of  $ \beta\omega_{\rm t}/2\pi\approx 60\,$kHz. As a direct consequence of the larger inter-ion spacing $d_{\rm m}$,  this tunneling is much smaller than in linear Paul traps. However, since there are more neighbours in the triangular lattice (i.e.  the value of $a_0$ is bigger than for linear chains),  the LR speed of propagation of spin correlations will not be much slower than the one found for  linear Paul traps. Let us now address the strength of the state-dependent dipole force. In the experiment~\cite{britton}, these forces are obtained from the gradient of a moving optical lattice formed by a couple of non-copropagating laser beams in a Raman configuration. 
For the  incident angles of these beams allowed by the experimental apparatus~\cite{britton}, these forces correspond to $g/2\pi\approx 0.6\,$kHz. By evaluating the LRB in Eq.~\eqref{lrb_ions_spin_boson2}, we find that the correlations can spread over a whole crystal of $N\sim100$-300 ions in a minimum time-scale of   1$\mu$s (see Fig.~\ref{fig_LRB_2d}). A clear advantage of Penning traps is that they can confine a   sufficiently-large number of ions, such  that the propagation of correlations becomes a real many-body problem very difficult to tackle even with the most sophisticated numerical methods. For this reason, the advent of a trapped-ion test of our LRB  would constitute a quantum simulation that overcomes the capabilities of classical computers.

Let us now consider another promising architecture, the so-called micro-fabricated surface traps~\cite{surface_traps}. By appropriate designing a planar electrode, it is possible to confine the ions above the electrode surface according to any desired geometry~\cite{schmied}. So far, in order to minimise the heating, the ions have been held sufficiently far away from the electrodes, such that typical ion-ion distances are much larger than in linear or  Penning traps (e.g. $d_{\rm m}\sim 40$-$50\,\mu$m for linear surface traps with $^9{\rm Be}^+$ ions~\cite{wineland_tunneling} or $^{40}{\rm Ca}^+$ ions~\cite{blatt_tunneling}). For such larger distances, the Coulomb couplings and thus the tunneling of vibrational excitations is reduced considerably. For instance, for $^9{\rm Be}^+$ crystals with $d_{\rm m}\approx 40\,\mu$m, and transverse trap frequency of  $\omega_{\rm t}/2\pi\sim10\,$MHz, we get $\beta\omega_{\rm t}/2\pi\approx 0.6\,$kHz. According to the LRB~\eqref{lrb_ions_spin_boson2}, we understand that 
the transport of correlations will be much slower in this case. For moderate state-dependent forces $g/2\pi\approx 0.4$kHz,  we find that the transport of correlations in the surface trap is  two orders of magnitude slower  with respect to the LRB of the linear chain in Fig.~\ref{fig_LRB_1d}\,{\bf (a)}. For the recent experiments~\cite{hensinger2}, where the fluorescence of a triangular crystal of  $^{171}{\rm Yb}^+$ ions in a surface trap has  been observed for the first time,  the estimated phonon tunneling  $\beta\omega_{\rm t}/2\pi\approx 0.03\,$kHz for trapping frequencies of $\omega_{\rm t}/2\pi\approx 3.3\,$MHz leads to a slower transport of correlations.

\begin{figure}
\centering
\includegraphics[width=.55\columnwidth]{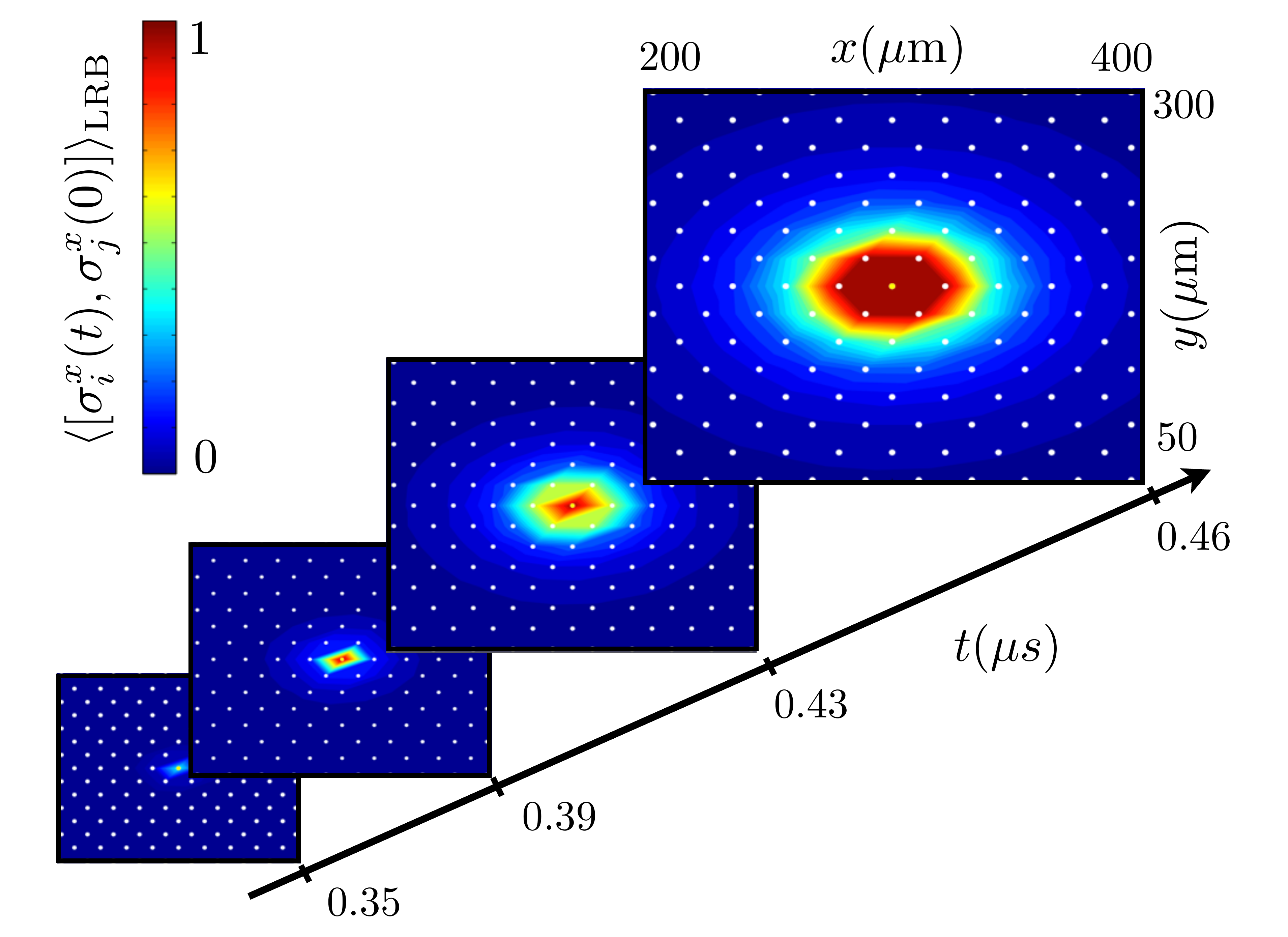}
\caption{ {\bf LRB for the spin correlations in a triangular ion crystal in a Penning trap: }. Evaluation of the spin-boson LRB in Eq.~\eqref{lrb_ions_spin_boson2} for a triangular crystal $^{9}{\rm Be}^+$ ions in triangular Penning trap (see the text for the specific parameters). We observe the evolution of a spin perturbation initially localised in the centre of the crystal, and spreading toward its boundary as the time evolves. }
\label{fig_LRB_2d}
\end{figure}

\subsubsection{Bounds for  perturbative interacting spin models}
\label{spin_spin}

There is a certain regime of the spin-boson lattice model~\eqref{sb_ions}, where the effect of the bosons as carriers of  spin correlations can be described neatly. This is the so-called far-detuned regime, where the spin-boson coupling~\eqref{G-H} is weak enough, such that  bosons can only be created/annihilated virtually. In this perturbative limit, one  traces out the bosons to obtain an effective spin interaction due to the virtual boson exchange between distant ions. To fix the notation, we  describe here such a derivation.

In order to trace out the bosons, it is more convenient to diagonalise first the harmonic crystal Hamiltonian~\eqref{tv_ham}. This can be done by the following canonical transformation 
\begin{equation}
\delta r_{i,\rm t}=\sum_n\sqrt{\frac{1}{{2m\omega_n}}}\mathcal{M}_{in}(a_n^{\dagger}+a_n),\hspace{2ex} p_{i,{\rm t}}=\ii\sum_n\sqrt{\frac{{m\omega_n}}{{2}}}\mathcal{M}_{in}(a_n^{\dagger}-a_n),
\end{equation}
where $a_n^{\dagger} (a_n)$ create(annihilate)  phonons in the crystal, and $\mathcal{M}_{in}$ are the elements of an orthogonal matrix that leads to the normal-mode frequencies of the crystal $\omega_n=\omega_{\rm t}(1+\beta\tilde{\mathcal{V}}_n)^{1/2}$. Here, $\tilde{\mathcal{V}}_n=\sum_{ij} \mathcal{M}_{in}\mathbb{\tilde{V}}_{ij}\mathcal{M}_{jn}$, and $\mathbb{\tilde{V}}_{ij}=1/|\vect{\tilde{r}}_i^0-\vect{\tilde{r}}_l^0|^3$ are the rescaled oscillator couplings, where  the equilibrium distances have been divided by the minimum distance of the crystal $\tilde{\vect{r}}_i^0=\vect{r}_i^0/d_m$. Hence, the harmonic crystal Hamiltonian~\eqref{tv_ham} becomes $H_{\rm b}=\sum_n\omega_na_n^{\dagger}a_n$. 

Let us now move to the interaction picture with respect to $H_0=\sum_iB^z\sigma_i^z+\sum_n\omega_na_n^{\dagger}a_n$. We assume that {\it (i)} the on-site spin terms~\eqref{spin_b_field} fulfil $\nu=\omega_0$, and $\varphi=0$, and $\Omega\ll\omega_0$, {\it (ii)} the state-dependent force~\eqref{sd_force} fulfils $\tilde{\nu}\approx\omega_n$, and $\tilde{\Omega}\gamma_n\ll\omega_n$, such that $\gamma_n=\gamma(\omega_{\rm t}/\omega_n)^{1/2}$. In this case, after a rotating-wave approximation, we can describe the interaction-picture Hamiltonian as
\begin{equation}
H(t)=\sum_ih\sigma_i^x+\sum_{in}\mathcal{F}_{in}\sigma_i^za_n^{\dagger}\ee^{\ii\tilde{\delta}_nt}+{\rm H.c.,}
\label{TIM_force}
\end{equation} 
where $h=\Omega/2$, $\mathcal{F}_{in}=\ii\tilde{\Omega}\gamma_n\ee^{\ii\tilde{\varphi}}\mathcal{M}_{in}/2$, and $\tilde{\delta}_n=\omega_n-\tilde{\nu}$ is the detuning of the state-dependent force. In the far-detuned regime $|\mathcal{F}_{in}|\ll\tilde{\delta}_n\ll\omega_n$, the force can only create/annihilate phonons virtually giving rise to an effective interaction between the spins. Assuming that the phonons are initially in a thermal state $\rho(t_0)=\ket{\psi_{\rm s}}\bra{\psi_{\rm s}}\otimes\rho_{\rm th}$, whereas the spins are in an arbitrary pure state $\ket{\psi_{\rm s}}$, it is possible to trace out the phonons by means of a canonical transformation~\cite{porras_ising} or via the Magnus expansion~\cite{magnus}. The latter leads to an effective time evolution operator for the spins that reads as follows
\begin{equation}
U_{\rm eff}(t)=\ee^{-\ii t H_{\rm eff}}+\mathcal{O}\big((g/\tilde{\delta}_{\rm t})^2(\bar{n}_{\rm t}+1/2)\big),
\label{magnus}
\end{equation}
where $g=\sqrt{2}\tilde{\Omega}\gamma$ is the strength of the spin-boson coupling, $\tilde{\delta}_{\rm t}$ is the detuning with respect to the center-of-mass  mode, and $\bar{n}_{\rm t}$ is its thermal occupation. Thus,  if the detuning is large enough, and the crystal is laser-cooled to sufficiently low temperatures  $(g/\tilde{\delta}_{\rm t})^2(\bar{n}_{\rm t}+1/2)\ll1$, the residual terms can be neglected. We thus obtain the  effective transverse-field Ising model
\begin{equation}
  H_{\mathrm{eff}} = \sum_{i\neq j} J_{ij} \sigma^z_i \sigma^z_j + \sum_i h \sigma^x_i,\hspace{2ex}J_{ij}=-\sum_{n}\frac{\mathcal{F}_{in}^*\mathcal{F}_{jn}}{\tilde{\delta}_n}.
  \label{tim}
\end{equation}

 For the transverse modes, where  $\beta = e_0^2/ m \omega_{\rm t}^2 d_{\rm m}^3\ll 1$, it is possible to show that the leading-order term for the spin-spin couplings decays algebraically with distance. In particular, if $\beta\ll 2\tilde{\delta}_{\rm t}/\omega_{\rm t}$, we obtain the  following a dipolar law
\begin{equation}
J_{ij}=\frac{J_0}{|{\tilde{\vect{r}}_i^0-\tilde{\vect{r}}_j^0}|^3},\hspace{2ex}J_0=\frac{1}{16}\bigg(\frac{g}{\tilde{\delta}_{\rm t}}\bigg)^2\beta\omega_{\rm t}
\end{equation}
 At this point, it is important to remark that the force is constrained to $(g/\tilde{\delta}_{\rm t})^2\ll1$, which follows from the need to neglect  residual spin-boson couplings in the evolution~\eqref{magnus}.  Therefore, the spin couplings $J_0\ll\beta\omega_{\rm t}$ are much smaller than the tunneling of phonons, which is consistent with the fact that the interactions are carried by the phonons via  perturbative virtual exchange. Although in this work we have focused on the regime of dipolar decaying  interactions $\beta\ll 2\tilde{\delta}_{\rm t}/\omega_{\rm t}$, let us also note that if   $\beta\omega_{\rm t}\approx2\tilde{\delta}_{\rm t}$, other algebraic decays can be found (e.g. Coulomb-like, monopole-dipole, etc). To achieve this regime, one may either reduce the detunings $\tilde{\delta}_{\rm t}$~\cite{britton}, or  change the vibrational bandwidth  $\beta\omega_{\rm t}$ modifying the axial trap frequency~\cite{monroe}. The latter method does not compromise the spin couplings, since the 
residual error $\mathcal{O}((g/\tilde{\delta}_{\rm t})^2)$ can be  fixed without decreasing the  forces.

In the dipolar regime, we can thus derive a LRB for the spin model following similar steps as in Sects.~\ref{LRB_spin_boson_sec} and~\ref{sec:lr-harmonic} (i.e. finding a Dyson-type recursion for the LR commutator,  bounding its norm, and resuming the expressions to infinite order). This derivation depends on the bound of the spin-spin couplings, and since  they share the same distance-dependence with the oscillator couplings $Q_{ij}$ (see Eq.~\eqref{bounds}), we require analogous bounds on the supremum norms
\begin{equation}
 \hspace{2ex}\norm{\boldsymbol{\sigma}_i}_{\infty}\leq S=1,\hspace{2ex} \norm{J_{ij}}_{\infty}=J_{ij} \le \frac{8J_0}{(1+d_{ij})^{3} }, \hspace{1ex}\forall i\neq j,
\end{equation}
where $d_{ij}$ is again the Euclidian distance of a perfect Bravais lattice that shares the geometry with the ion crystal.
We can then establish the following LRB for the effective spin model
\begin{equation}
\norm{[\vect{\sigma}_i(t), \sigma^{\phi}_j(0)]}_{\infty} \leq {2 \over a_0 (1 + d_{ij})^3} \bigg( e^{\tilde{\alpha}\left(\frac{g}{\tilde{\delta}_{\rm t}}\right)^2(\beta\omega_{\rm t})t} -1 \bigg),\hspace{1ex}\tilde{\alpha}=a_0
\label{lrb_ions_spins}
\end{equation}
where $a_0$ is again defined by the maximisation of the convolution~\eqref{maximization}. We note that this bound coincides with the formal result of~\cite{nachtergaele} applied to our case. Let us emphasise that 
the parameter dependence of this spin-LRB resembles the spin-boson-LRB found in Eq.~\eqref{lrb_ions_spin_boson2}. There are, however, two main differences: {\it (i)} As the bosons have been traced out by a sort of adiabatic elimination, their propagation (i.e. first exponential in Eq.~\eqref{lrb_ions_spin_boson2}) does not appear in the spin-LRB. {\it (ii)} The term in brackets, which accounts for the spin-spin coupling by virtual boson exchange, scales with $(g/\tilde{\delta}_{\rm t})^2$ as opposed to $(g/\beta\omega_{\rm t})^2$ for the spin-boson-LRB~\eqref{lrb_ions_spin_boson2}. Let us now discuss  realistic values for different setups.

{\it (i) Ion chain in a linear Paul trap.--}  We consider $^{25}{\rm Mg}^+$ ions confined in a trap with the same parameters as discussed for the spin-boson LRB. The only parameter that we have to modify is the strength of the state-dependent force to fulfil the far-detuned-regime condition, such that  the effective spin model is an accurate description. Let us first fix a large detuning $\tilde{\delta}
_{\rm t}/2\pi\approx 0.5\,$MHz, which fulfils $\tilde{\delta}_{\rm t}\ll\omega_{\rm t}$.  To reach the far-detuned regime $g\ll\tilde{\delta}_{\rm t}$, we choose $g/2\pi\approx 50\,$kHz. By substituting the previously-found value $a_0=2.9$, and considering again a chain of $N=30$ ions, the corresponding spin-LRB~\eqref{lrb_ions_spins} leads to the correlation transport  displayed in Fig.~\ref{fig_LRB_1d}\,{\bf (b)}. In contrast to the speed of correlations predicted by the spin-boson LRB (see Fig.~\ref{fig_LRB_1d}\,{\bf (a)}), we find that the LR bound for the effective spin model predicts a much slower spread of correlations in the 0.1\,ms range.

\vspace{0.5ex}
{\it (ii) Triangular lattice in a Penning or surface trap.--} Let us now discuss the orders of magnitude for the propagation speed in the far-detuned regime of Penning traps and surface traps. For  $^{9}{\rm Be}^+$ ions in Penning traps, we fix again the detuning  $\tilde{\delta}_{\rm t}/2\pi\approx 80\,$kHz. For the weak forces attained in the experiment  $g/2\pi\approx 0.6\,$kHz~\cite{britton}, and recalling that $ \beta\omega_{\rm t}/2\pi\approx 60\,$kHz, the LRB predicts a propagation of spin correlations  in the milliseconds range. By allowing for larger incident angles of the laser beams responsible for the force, it is expected to achieve stronger forces $g/2\pi\approx$4\,kHz~\cite{britton} that would allow for LRB in the 0.1\,ms range. Achieving such propagation speeds is important, as other sources of noise (e.g. magnetic field fluctuations) lead to decoherence times in  1-10\,ms timescales. Let us now address a  surface trap loaded with $^{9}{\rm Be}^+$ ($^{171}{\rm Yb}^+$ ) ions 
forming a triangular lattice. Let us recall that the ion spacing in this case was much larger, such that $\beta\omega_{\rm t}/2\pi\approx 0.6\,$kHz ($\beta\omega_{\rm t}/2\pi\approx 0.03\,$kHz). Considering the same detunings and forces as for the Penning trap, this leads to a propagation in    $0.1$-$1\,$s (1-10\,s), far too slow with respect to existing sources of noise.  According to this discussion, we can conclude that the transport of spin correlations in the far-detuned regime can be sometimes hindered by existing sources of noise in the experiment. From a pragmatic point of view, it would be very interesting to study how to approach the faster spin-boson LRB~\eqref{lrb_ions_spin_boson2} experimentally.

\subsubsection{Bounds for  impulsive spin-boson models}

 The above evaluation of the bounds  has shown that the speed of propagation of spin correlations in the far-detuned regime is at least two orders of magnitude slower than the prediction of the spin-boson LRB (compare Figs.~\ref{fig_LRB_1d}\,{\bf (a)} and {\bf (b)}).  Interestingly, by abandoning the far-detuned regime such that  bosons cannot be eliminated from the  dynamics, our new LRB~\eqref{lrb_ions_spin_boson2} predicts that there is plenty of room for enhancement of the propagation speed. In fact, it seems possible that the spin correlations spread with the maximum possible speed: the speed  of the bare bosons propagating in the lattice. However, the LR theory does not tell us how to achieve this bound in practice, i.e. it does not specify the particular spin-boson coupling or the time-modulation of the Hamiltonian parameters that would allow us to reach the aforementioned LRB~\eqref{lrb_ions_spin_boson2}. Finding the optimal regime of the spin-boson lattice model poses a  many-body problem much more 
difficult to tackle, even numerically, than the effective spin  model (see e.g.~\cite{ent_growth}). Hence, the possibility of exploring the LRB experimentally would be an instance of a quantum simulator that overcomes the power of the most sophisticated algorithms with classical computers.

As a guiding principle, we now study a simplified scenario that suggests that the optimal propagation speed~\eqref{lrb_ions_spin_boson2} could also be achieved in the truly many-body problem~\eqref{full-H}.  Let us consider the trapped-ion Hamiltonian~\eqref{TIM_force} for $h=0$. In this limit,  $\sigma_i^z(t)=\sigma_i^z(t_0)$ is a conserved quantity, and the dynamics of the spin-boson lattice model can be  integrated exactly.  
By using the free boson propagator $W_{ij}(\tau_1,\tau_2)$ in Eq.~\eqref{eq-source}, it is possible to find the following bound for the LR commutator
\begin{equation}
\norm{[\sigma_i^x(t_{\rm f}),\sigma_j^x(t_0)]}_\infty\leq 8\sin\left(\int_{t_0}^{t_{\rm f}}\!\!\!\dd\tau_1\int_{t_0}^{\tau_1}\!\!\!\dd\tau_2F_{z,i}(\tau_1)W_{ij}^{xp}(\tau_1,\tau_2)F_{z,j}(\tau_2)\right),
\end{equation} 
which involves the state-dependent forces acting on the two ions  $F_{z,i}(\tau_1),F_{z,j}(\tau_2)$ at different time-ordered instants  $\tau_1>\tau_2$. Let us remark that this expression does not require  summing the Dyson series to infinite order as we did for the spin-boson LRB~\eqref{lrb_ions_spin_boson2}. It rather follows from the exact integrability of the dynamics, and thus serves as a test-bed for the validity of Eq.~\eqref{lrb_ions_spin_boson2}. We will focus on the impulsive regime, where the forces act locally on the distant ions for a very short interval of time $\delta t\sim1/g$, such that $\delta t\ll(\beta\omega_{\rm t})^{-1}$. Under this constraint, the phonons do not propagate
during the time where the pulsed forces are active, and we can approximate the forces as Dirac delta functions
\begin{equation}
F_{z,i}(\tau_1)=\theta_i\delta(\tau_1-t_{\rm f}),\hspace{1ex} F_{z,j}(\tau_2)=\theta_j\delta(\tau_2-t_{\rm 0}),\hspace{2ex}\theta_l=\int_{t_0}^{ t_{\rm f}}\!\!\!\dd \tau F_{z,l}(\tau).
\end{equation}
Here, the pulse area $\theta_l$ is related to the number of local vibrational excitations created by each force (i.e. $\bar{n}_l=|\theta_l|^2\sim (g\delta t)^2$). In this impulsive regime, we obtain
\begin{equation}
\norm{[\sigma_i^x(t_{\rm f}),\sigma_j^x(t_0)]}_\infty\leq 8\sin\left( W_{ij}^{xp}(t_{\rm f},t_0)\theta_i\theta_j\right)\leq 8 |W_{ij}^{xp}(t_{\rm f},t_0)|\times|\theta_i\theta_j|.
\label{LRB_impulsive2}
\end{equation} 
We have thus obtained that the propagation of correlations in this impulsive regime is given by two contributions: the bare propagation of the phonons, and a term that depends on the efficiency of the spin-phonon coupling in correlating spins and phonons. This is exactly the form of the more general spin-boson LRB~\eqref{lrb_ions_spin_boson2}. This result is also intuitively correct, as we are {\it (i)} using a fast  force to excite the phonons correlating them to the spin state at $t=t_0$, {\it (ii)} letting the bosons evolve under no additional force for $t\in(t_0,t_{\rm f})$, and {\it (iii)} performing another fast force to correlate the propagated phonons to a distant ion at $t=t_{\rm f}$.

Let us now  go back to the state-dependent force of strength $g=\sqrt{2}\tilde{\Omega}\gamma$ in Eq.~\eqref{sd_force}, and address the possibility of reaching the impulsive regime $g\sim(\delta t)^{-1}\gg\beta\omega_{\rm t}$ in ion-trap experiments. As argued below this equation, to achieve this force, the frequency of the radiation must be tuned $\tilde{\nu}\approx\omega_{\rm t}$, and its strength  constrained to $\tilde{\Omega}\ll\omega_{\rm t}$ as we want to make  the gradient of the radiation dominant with respect to other sidebands. For Lamb-Dicke parameters $\gamma\sim 0.1$, this poses a constraint on the achievable forces $g\lesssim 10^{-2}\omega_{\rm t}$. Moreover, considering the stiffness parameters of the above experimental realisations $\beta(^{25}{\rm Mg}^+,{\rm Linear})\approx0.09$, $\beta(^{9}{\rm Be}^+,{\rm Penning})\approx0.08$, $\beta(^{9}{\rm Be}^+,{\rm Surface})\approx 0.06\cdot 10^{-3}$, it is clear that the impulsive regime $g\gg\beta\omega_{\rm t}$ could only be attained for 
surface traps using this implementation of the forces. We  now discuss two possible alternatives to reach the impulsive regime:

 {\it (i)} By concatenating pairs of short resonant laser pulses coming from different directions, it is possible to implement much stronger state-dependent forces in the $\sigma_x$-basis without the requirement of resolving the sidebands~\cite{fast_gates}. As shown in recent experiments~\cite{fast_kicks}, this allows for very fast state-dependent forces $\delta t\approx 3\,$ns that create $\bar{n}\approx 10$ phonons. Therefore, one would obtain very strong and fast forces $g\sim\sqrt{\bar{n}}/\delta t\approx2\pi\times 170\,$ MHz, which would clearly fulfil the impulsive-regime constraint for any of the above realisations. However, one should also note the technical overhead of this method, as it requires the use of pulsed trains of ultrafast picosecond laser pulses~\cite{fast_kicks}. 
 
 {\it (ii)} We now discuss an alternative without these experimental requirements which, although not allowing for such strong forces, can still reach the impulsive regime for $^{9}{\rm Be}^+$ in Penning traps.  The main message is that one can alleviate the condition of the resolution of the sidebands $\tilde{\Omega}\ll\omega_{\rm t}$ to $\tilde{\Omega}\gamma\ll4\omega_{\rm t}$. Under this condition, in addition to the gradient~\eqref{TIM_force}, we should also consider the homogeneous terms as they can no longer be neglected. On the contrary, all the higher sidebands of the spin-phonon coupling can be neglected, and for $h=0$ and $\tilde{\nu}\approx\omega_{\rm t}$, we obtain 
 \begin{equation}
H(t)=\sum_i\frac{\tilde{\Omega}}{2}\sigma_i^z\ee^{-\ii{\tilde{\nu}}t}+\sum_{in}\mathcal{F}_{in}\sigma_i^za_n^{\dagger}\ee^{\ii\tilde{\delta}_nt}+{\rm H.c.}.
\end{equation}
Another factor that typically limits the  strength of the state-dependent forces in experiments is the compensation of ac-Stark shifts whereby photons are absorbed and reemitted into the same laser beam~\cite{britton}. It is important to compensate such shifts when the forces are applied for a long period of time, as fluctuations in the laser intensities would lead to decoherence. However, for the short pulses required in the impulsive regime, these ac-Stark shifts need not be compensated as they can be refocused by a simple spin echo provided that the laser intensities do not fluctuate during $\delta t\sim g^{-1}$. We thus incorporate possible ac-Stark shifts to the spin-boson Hamiltonian
 \begin{equation}
H(t)=\sum_{i}\frac{1}{4}\Delta\omega_{{\rm ac}}\sigma_i^z+\sum_i\frac{\tilde{\Omega}}{2}\sigma_i^z\ee^{-\ii{\tilde{\nu}}t}+\sum_{in}\mathcal{F}_{in}\sigma_i^za_n^{\dagger}\ee^{\ii\tilde{\delta}_nt}+{\rm H.c.}.
\end{equation}
The problem can still be integrated exactly, leading to an evolution operator $U(\delta t)=\ee^{-\ii\sum_i(\frac{1}{2}\Delta\omega_{\rm ac}\delta t+\frac{\tilde{\Omega}}{\tilde{\nu}}\sin(\tilde{\nu}\delta t))\sigma_i^z}U_{\textsf{SBL}}(\delta t)$, where $U_{\textsf{SBL}}(t)$ is the evolution operator leading to the LRB in the impulsive regime~\eqref{LRB_impulsive2}. We first impose that $\nu \delta t=2\pi n$, where $n\in\mathbb{Z}$. Additionally, at the middle of the evolution we apply a spin-echo pulse consisting of a $\pi$-pulse $\sigma_i^z\to-\sigma_i^z$, and $\tilde{\Omega}\to-\tilde{\Omega}$. The $\pi$-pulse  is routinely achieved in trapped-ion experiments by driving the carrier transition~\cite{haeffner_review2}, whereas the inversion of the Rabi frequency can be achieved by controlling the laser phase~\cite{lm_gate}. In this way, $U(\delta t)=U_{\rm SBL}(\delta t)$, and we can overcome the effects of the spurious terms. In this new regime, taking into account the parameters of $^{9}{\rm Be}^+$, and the 
larger incident angles planned in the experiment~\cite{britton}, we find that the forces can be as large as $g/2\pi\approx0.3\,$MHz, such that $\beta\omega_{\rm t}/g\approx 0.2$ and we achieve the desired impulsive. Note that the pulsed forces are applied for time intervals of $\delta t\sim$0.1-1$\,\mu$s, which is considerably shorter in comparison to the propagation of the spin correlations.

Let us close this section by reminding that this impulsive regime should serve as a guiding principle to test experimentally how the LRB~\eqref{lrb_ions_spin_boson2} can be attained. However, we should keep in mind  that the interesting many-body problem would be the one where the forces are non-perturbative and also non-impulsive. 

\subsection{Probing the Lieb-Robinson bound through fluorescence measurements}
\label{sect_measurment}

In this section, we describe how to probe the LRB in a trapped-ion experiment. We will exploit the high accuracies in controlling and measuring the quantum state of a collection of trapped ions~\cite{haeffner_review2}.  Let us emphasise that the experimental scheme, which has been depicted in Fig.~\ref{fig_LRB_measurement2}, consists of a sequence of operations that are standard  in several trapped-ion laboratories dedicated to quantum-information processing. This sequence can be divided in three steps:

\begin{figure}
\centering
\includegraphics[width=.7\columnwidth]{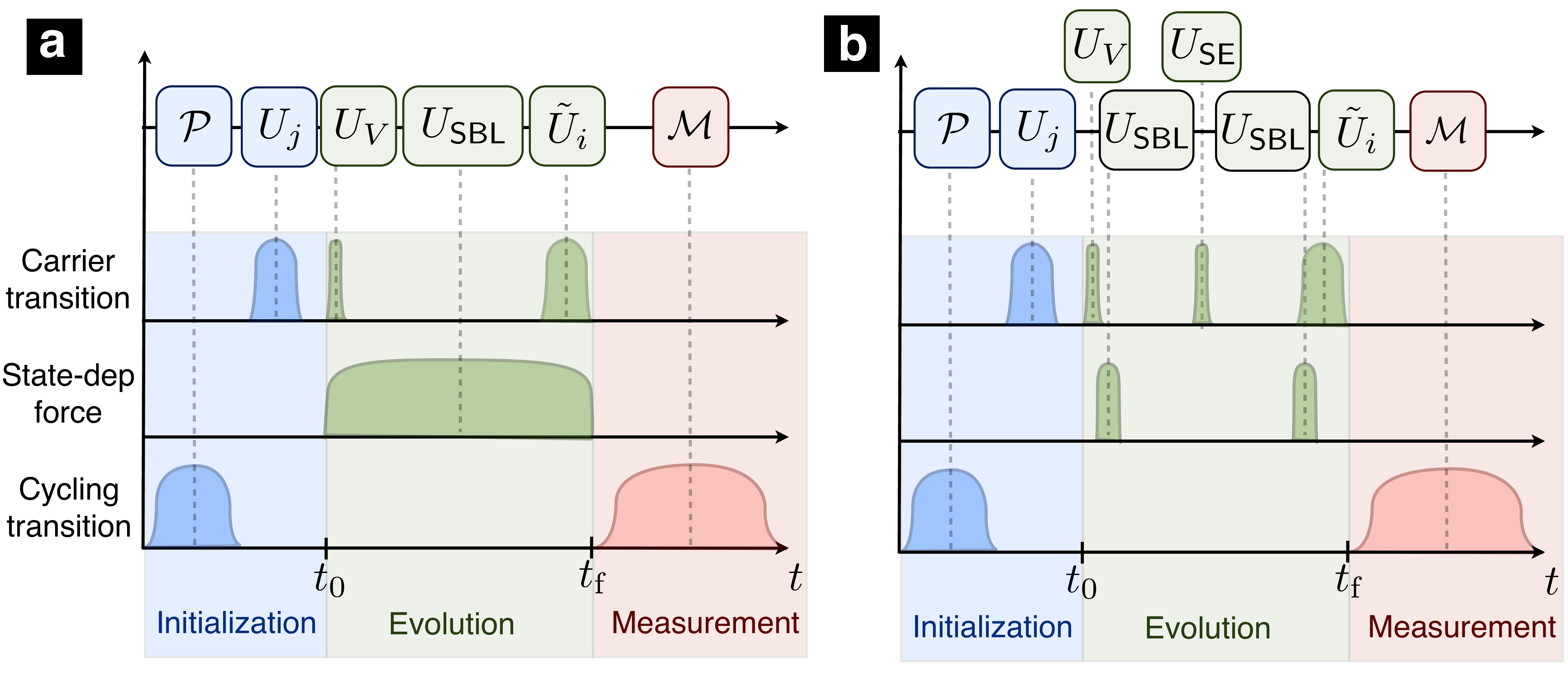}
\caption{ {\bf Experimental sequence to test the LRB: }  {\bf (a)} Always-on, and {\bf (b)} pulsed spin-phonon forces. We represent  the {\it initialisation} step in blue, which consists of laser cooling followed by optical pumping $\mathcal{P}$, which leads to $\ket{{\downarrow\cdots\downarrow}}\bra{{\downarrow\cdots\downarrow}}\otimes\rho_{\rm th}$, where $\rho_{\rm th}$ is a thermal state of the phonons after Doppler cooling. We then apply a $\pi/2$-pulse $U_{j}={\rm exp}\{\ii\frac{\pi}{2}\sigma_j^y\}$ by driving the carrier transition~\cite{haeffner_review2}. In the {\it measurement} step in red, one collects  the state-dependent fluorescence $\mathcal{M}$ during a continuous driving of the cycling transition~\cite{haeffner_review2}. At the beginning of the {\it evolution} step $t=t_0$, we  apply the unitary $U_{V}$ associated to the impulsive perturbation $V(t)$ described in the main text. This is followed by the actual evolution under the state-dependent forces: {\bf (a)} in the always-on regime, the 
forces should be switched on continuously during the evolution, or {\bf (b)} in the impulsive regime, we apply to pulsed state-dependent forces during a short time interval. Additionally, at the middle of the evolution, we apply the spin-echo sequences $U_{\textsf{SE}}$ consisting of $\sigma_i^z\to-\sigma_i^z$ and $\tilde{\Omega}\to-\tilde{\Omega}$. Before measuring, we apply   another $\pi/2$-pulse $\tilde{U}_i={\rm exp}\{-\ii\frac{\pi}{2}\sigma_i^y\}$. }
\label{fig_LRB_measurement2}
\end{figure}

 {\it (i)} The first step is the {\it initialisation}, namely   to prepare a localised spin excitation in a certain region of the ion crystal at $t=t_0$. Considering the  trapped-ion realisation of the spin-boson coupling~\eqref{G-H}, and the effective Ising interaction in the perturbative limit~\eqref{tim}, we will study the following initial state $\rho(t_0)=\ket{\psi_{\rm s}}\bra{\psi_{\rm s}}\otimes\rho_{\rm th}$, where $\rho_{\rm th}$ is the thermal state of the vibrational excitations after laser cooling, and $\ket{\psi_{\rm s}}=U_{j}\ket{\downarrow\cdots\downarrow}$ is obtained by optical pumping $\mathcal{P}$ to a state where all spins pointing down $\ket{\downarrow\cdots\downarrow}$, and subsequently implementing a $\pi/2$-pulse  at a certain ion $j$, namely $U_{j}={\rm exp}\{\ii\frac{\pi}{2}\sigma_j^y\}$. Ideally, $\ket{\psi_{\rm s}}=\ket{\downarrow\cdots\downarrow+_{j}\downarrow\cdots\downarrow}$, where $\ket{+_{j}}=(\ket{\uparrow_{j}}+\ket{\downarrow_{j}})/\sqrt{2}$ is the spin excitation. 
However, we remark that the LRB would also apply if the initial perturbation is delocalised around $j$, as far as it does not have an overlap with the distant lattice site $i$ where the measurement  takes place. Therefore, the experiment does not  require individual addressability. Moreover, we also emphasise that  laser cooling to the vibrational ground-state is not required, as the general LRB~\eqref{lrb_ions_spin_boson2} is valid for any temperature of the ions (provided that the crystal is stable, and only small excursions from the equilibrium positions take place). This is a clear advantage with respect to the effective spin models~\eqref{tim}, which are obtained by tracing out the vibrational excitations, and whose validity relies on minimising residual spin-phonon couplings. This requires either cooling closer to the vibrational ground-state, or using larger detunings such that the couplings become weaker, and other sources of noise may start contributing. In our case,
Doppler cooling to modest temperatures (e.g. $\bar{n}_i={\rm Tr}\{a_i^\dagger a_i\rho_{\rm th}\}\sim$10-20) will suffice to test the validity of the LRB.  Finally, note also that  unitaries analogous to $U_{j}$ correspond to  single-qubit gates for quantum computation, which have been accomplished with very high fidelities~\cite{haeffner_review2}. Due to all these properties, the initialisation step can be achieved with accuracies above $99\%$.

 {\it (ii)} After state preparation, the following step in Figs.~\ref{fig_LRB_measurement2}{\bf (a)} and {\bf (b)} would be  the {\it evolution} for $t\in[t_0,t_{\rm f})$, where we  switch on the spin-boson lattice Hamiltonian $H_{\textsf{SBL}}(t)$~\eqref{sb_ions} continuously (Fig.~\ref{fig_LRB_measurement2}{\bf (a)}) or in a couple of short uses (Fig.~\ref{fig_LRB_measurement2}{\bf (b)}). We let the spin excitation propagate in time $\rho(t_{\rm f})=U_{\textsf{total}}(t_{\rm f},t_0)\rho(t_0)U_{\textsf{total}}^{\dagger}(t_{\rm f},t_0)$, where $U_{\textsf{total}}(t_{\rm f},t_0)=U_{\textsf{SBL}}(t_{\rm f}, t_0)$ in the continuous regime of Fig.~\ref{fig_LRB_measurement2}{\bf (a)}, and $U_{\textsf{total}}(t_{\rm f},t_0)=U_{\textsf{SBL}}(t_{\rm f}, \frac{1}{2}(t_{\rm f}-t_0))U_{\textsf{SE}}U_{\textsf{SBL}}(\frac{1}{2}(t_{\rm f}-t_0),t_0)$ in the pulsed regime of Fig.~\ref{fig_LRB_measurement2}{\bf (b)}. Here, we have introduced the evolution operator under the spin-boson lattice model~\eqref{sb_ions}, namely
\begin{equation}
U_{\textsf{SBL}}(t_{\rm f}, t_0)=\mathcal{T}\left\{\ee^{-\ii\int_{t_0}^{t_{\rm f}}\dd\tau H_{\textsf{SBL}}(\tau)}\right\},
\label{sb_evolution}
\end{equation}
and the corresponding spin echo $U_{\textsf{SE}}$ that acts at the middle of the evolution.
  Let us note that the use of state-dependent dipole forces, such as the force  in the $z$-basis~\eqref{sb_ions} or in $x,y$-bases, has become a frequent tool in different laboratories~\cite{laser_state_dep_forces}. Such forces underlie a wide variety of quantum-information experiments, such as the creation of Schr\"odinger cat states with single ions,  conditional phase gates for quantum computing with two ions, or quantum simulations of magnetic interactions with several ions. Therefore, we expect that the evolution  step can also be conducted with very high accuracies.

 {\it (iii)} Once the state of the system has evolved in time $\rho(t_0)\to\rho(t_{\rm f})$, the {\it measurement} step of the protocol starts (Fig.~\ref{fig_LRB_measurement2}). In order to test the LRB~\eqref{lrb_ions_spin_boson2}, we need to measure the retarded correlation function  $C_{\sigma_i^{x},\sigma_j^{x}}(t_{\rm f}-t_0)=\langle [\sigma_i^{x}(t_{\rm f}),\sigma_j^{x}(t_0)]\rangle$. However, the usual trapped-ion 
measurements $\mathcal{M}$ based on state-selective fluorescence only allow for measurements of single-time observables (e.g. $\langle \sigma_i^{z}(t)\rangle, \langle \sigma_i^{z}(t)\sigma_j^{z}(t)\rangle$)~\cite{haeffner_review2}. In the following, we describe a modification of these schemes for the measurement of the above retarded correlation function. The main idea is to encode the information of the retarded correlator in the measurement of a single-time observable by means of a certain perturbation applied during the evolution step (i.e. a  linear-response-type scheme~\cite{lin_response}). To maintain the generality, let $A_i$ be the single-spin observable that can be measured at $t=t_{\rm f}$. At $t=t_0$, we let the system evolve under a perturbed spin-boson lattice Hamiltonian  $H(t)=H_{\textsf{SBL}}(t)+V(t)$, where $V(t)=\lambda_{\rm B} B_j\delta(t-t_0)$ with $\lambda_{\rm B}\ll 1$ is a dimensionless perturbative parameter, $B_j$ is a certain  operator localised around $j$, and $\delta(t-t_0)$ is the 
Dirac delta function. At  $t=t_{\rm f}$, we switch off the perturbed spin-boson lattice Hamiltonian, and perform an additional unitary operator $\tilde{U}_i$ localised around the site $i$ (see Fig.~\ref{fig_LRB_measurement2}), consisting of single-spin rotations (i.e. single-qubit gates). Using the  interaction-picture formalism, the total time-evolution operator can be thus written as follows
\begin{equation}
U=\tilde{U}_iU_{\textsf{total}}(t_{\rm f},t_0)U_{V}(t_{\rm f},t_0),\hspace{2ex}U_{V}(t_{\rm f},t_0)= \mathcal{T}\left\{\ee^{-\ii\int_{t_0}^{t_{\rm f}}\dd\tau \hat{V}(\tau)}\right\},
\end{equation}
where $\hat{V}(\tau)=U^{\dagger}_{\textsf{total}}(\tau,t_0){V}(\tau)U_{\textsf{total}}(\tau,t_0)$.
Due to the impulsive and perturbative nature of the perturbation, we can approximate this evolution operator as $U\approx\tilde{U}_iU_{\textsf{total}}(t_{\rm f},t_0)(\openone-\ii\hat{V}(t_0))$. Finally, the measurement of the observable gives us $\langle A_i(t_{\rm f})\rangle_{\rm pert} ={\rm Tr}\{(\tilde{U}_iU_{\textsf{total}}(t_{\rm f},t_0)(\openone-\ii\lambda_{\rm B}B_j(t_0)))^{\dagger}A_i\tilde{U}_iU_{\textsf{total}}(t_{\rm f},t_0)(\openone-\ii\lambda_{\rm B}B_j(t_0))\rho(t_0)\}$. To linear order in the perturbation strength, i.e. linear-response theory, we find
\begin{equation}
\langle A_i(t_{\rm f})\rangle_{\rm pert}=\langle \tilde{A}_i(t_{\rm f})\rangle_{\rm unpert}-\ii\lambda_{\rm B}\langle[\tilde{A}_i(t_{\rm f}),B_j(t_0)]\rangle_{\rm unpert},
\end{equation}
where we have defined $\tilde{A}_i=\tilde{U}^{\dagger}_i A_i \tilde{U}_i$, and the subindex $\langle\cdot\rangle_{\rm unpert}$ refers to the expectation value for the time-evolved state with respect to the unperturbed Hamiltonian, namely the spin-boson lattice model $U_{\textsf{total}}(t_{\rm f},t_0)$ in the continuous or pulsed regimes. Therefore, by letting the system evolve with and without the perturbation, we can measure $f(\lambda_{\rm B})=\langle A_i(t_{\rm f})\rangle_{\rm pert}-\langle \tilde{A}_i(t_{\rm f})\rangle_{\rm unpert}$, and thus estimate the retarded correlator. To be more precise, $\dd f/\dd\lambda_{\rm B}|_{\lambda_{\rm B}=0}=-\ii\langle[\tilde{A}_i(t_{\rm f}),B_j(t_0)]$, so we would need to modify the perturbative parameter $\lambda_{\rm B}$, such that we can estimate its derivative for very weak couplings. We note that the use of measurement unitaries $\tilde{U}_i$ has  been already demonstrated in  the measurement of single-time observables in different basis (e.g. $\langle \sigma_i^{\
\alpha}(t)\rangle, \langle \sigma_i^{\alpha}(t)\sigma_j^{\beta}(t)\rangle$) for state tomography~\cite{state_tomography2}, or to recover the position operator of a trapped ion~\cite{position_measurement}. By including a perturbation  at $t=t_0$,  we  get access to two-time observables, and in particular to the desired information about the LR commutator.

Let us now apply this scheme to the dynamics under study. In this case, the state-dependent fluorescence allows us to measure $A_i =\sigma_i^z$. Typical fidelities for this type of measurements are above $99\%$ for photon-collection times on the millisecond range~\cite{haeffner_review2}. Although the spin correlation transport occurs on a much faster time-scale, the fact that we  switch off the spin-phonon coupling at $t=t_{\rm f}$ (Fig.~\ref{fig_LRB_measurement2}) implies that the spin-populations will be frozen for $t>t_{\rm f}$. Thus, this scheme allows for the required photon-collection times without compromising the information about the transport of correlations. The perturbation that must be applied to recover the desired correlator is $B_j=\sigma_j^x$, which can be achieved by driving the so-called carrier transition of the ions, such that $\lambda_{\rm B}=\Omega\delta t /2$. We can reach the perturbative  regime by simply driving the carrier with a sufficiently small intensity. Therefore, the Rabi 
frequency $\Omega$ must be much smaller than any other coupling strength in the problem $\Omega\ll \{g,\beta\omega_{\rm t}\}$. Moreover, the impulsive regime will be a good approximation when the time during which the perturbation is applied, $\delta t$, is  much smaller than any other time-scale of the problem $\delta t\ll\{g^{-1},(\beta\omega_{\rm t})^{-1}\}$. Finally, the measurement unitary corresponds to $\tilde{U}_i=\ee^{-\ii\frac{\pi}{2}\sigma_i^y}$, which leads to the desired correlator $C_{\sigma_i^{x},\sigma_j^{x}}(t_{\rm f}-t_0)$ encoded in the resonance fluorescence of the ion
\begin{equation}
\langle \sigma_i^z(t_{\rm f})\rangle_{\rm pert}=\langle \sigma_i^x(t_{\rm f})\rangle_{\rm unpert}-\ii\lambda_{\rm B}\langle[\sigma^x_i(t_{\rm f}),\sigma_j^x(t_0)]\rangle_{\rm unpert}.
\end{equation}
As announced previously, by measuring a single-time observable in the presence of a perturbation, we can recover the retarded correlator and test the validity of the LRB. At this point, it is worth commenting on the following points. 
First, let us note that the unitary $\tilde{U}_i$ could also be delocalised around the site $j$, such that individual addressability is not required. Second, we remark that other choices of $B_j,\tilde{U}_i$, which are equally accessible in an experiment, would allow us to estimate any other correlator $\langle [\sigma^{\alpha}_i(t), \sigma^{\beta}_j(t_0)]\rangle$, which might be important when the state-dependent forces act in a different basis, or if we combine them to produce Heisenberg-type Hamiltonians. Moreover, the use of state-dependent forces in $\tilde{U}_i$ can also allow for measurements of the LR commutators for the free bosonic lattice to test the harmonic LRB~\eqref{w_bound}.

\end{widetext}


\begin{thebibliography}{}
\bibitem{comment_lie_algebra}
 For concreteness, we use a $d$-dimensional representation of $\mathfrak{su}(2)$, $ [S_i^\alpha, S_j^\beta] = i \delta_{i,j} \sum_{\gamma} f^{\alpha\beta\gamma} S_i^\gamma$, where $f^{\alpha\beta\gamma} = 2\epsilon^{\alpha\beta\gamma}$ is proportional to the Levi-Civita symbol and $m=3$. However, our LRBs apply to any Lie algebra with fully antisymmetric $f^{\alpha\beta\gamma}$.
\bibitem{comment_ions}
The spin states may correspond to a pair of levels from the hyperfine ground-state manifold of $^9{\rm Be}^+,^{25}{\rm Mg}^+, ^{171}{\rm Yb}^+$, a pair of Zeeman sub-levels of $^{88}{\rm Sr}^+,$ or a ground-state and an excited metastable state of $^{40}{\rm Ca}^+$.
\bibitem{comment_temperature}
    The effective Ising model is only valid for cold-enough crystals: residual spin-phonon couplings can only be neglected if $(g/{\delta}_{t})^2(\bar{n}_{t})\ll 1$, where $\bar{n}_{t}$ is the average number of phonons in the transverse modes after laser cooling. Hence, the LRB of Eq.~\eqref{lrb_ions_spin_spin} is only strictly correct when the crystal is at almost zero temperature and for weak effective interactions. Our LRB of Eq.~\eqref{lrb_ions_spin_boson} is, however, rigorous for any force and temperature.
\bibitem{addressability} We remark that the LRB would also apply if the excitation is not perfectly localized at the $j$-th ion, provided it does not overlap with the distant $i$-th ion where we make the measurement. The same occurs for unitaries $U_V$ and $\tilde{U}_i$ in Fig.\ \ref{fig_LRB_measurement} so that we do not require perfect single-ion addressability with the lasers.
\bibitem{comm_position}
In the spirit of the measurement of $\langle \vect{R}^T_i(t)\rangle$ in Ref.~\cite{position}, if state-dependent forces are applied, it is  possible to recover correlation functions for the bosonic operators $\langle \vect{R}^T_i(t)\vect{R}_j(0)\rangle$.
\end{thebibliography}
\end{document}